\documentclass{aa}

\usepackage{amsmath,amssymb,amsthm}
\usepackage{mathabx,mathrsfs}
\usepackage{natbib}
\usepackage{graphicx}
\usepackage{svg}
\usepackage{txfonts}
\usepackage{atbegshi}
\usepackage[para,online,flushleft]{threeparttable}
\bibpunct{(}{)}{;}{a}{}{,}

\usepackage{hyperref}
\begin{document}

    \title{
    Comparative analysis of atmospheric parameters from high-resolution spectroscopic sky surveys: APOGEE, GALAH, Gaia-ESO}
    \author{Viola~Heged\H{u}s\inst{1,2}
    \and
    Szabolcs~M{\'e}sz{\'a}ros\inst{2,3}
    \and
    Paula~Jofr{\'e}\inst{4}
    \and
    Guy~S.~Stringfellow\inst{5}
    \and
    Diane~Feuillet\inst{6}
    \and
    Domingo~An{\'i}bal ~Garc\'{\i}a-Hern{\'a}ndez\inst{7}
    \and
    Christian~Nitschelm\inst{8}
    \and
    Olga~Zamora\inst{7}}

\institute{ELTE Eötvös Loránd University, Doctoral School of Physics, Budapest, 1117           Budapest Pázmány Péter sétány 1/A, Hungary
    \and
    MTA-ELTE Lend{\"u}let "Momentum" Milky Way Research Group, Hungary
    \and
    ELTE E\"otv\"os Lor\'and University, Gothard Astrophysical Observatory, 9700 Szombathely, Szent Imre H. st. 112, Hungary
    \and 
    N{\'u}cleo Milenio ERIS \& N{\'u}cleo de Astronom{\'i}a, Facultad de Ingenier{\'i}a y Ciencias, Universidad Diego Portales, Santiago de Chile
    \and
    Center for Astrophysics and Space Astronomy,  University of Colorado, 389 UCB, Boulder, CO 80309-0389, USA
    \and
    Lund Observatory, Department of Astronomy and Theoretical Physics, Lund University, Box 43, SE-22100 Lund, Sweden
    \and
    Instituto de Astrof{\'i}sica de Canarias, 38205 La Laguna, Tenerife, Spain
    \and
    Centro de Astronom{\'i}a (CITEVA), Universidad de Antofagasta, Avenida Angamos 601, Antofagasta 1270300, Chile}

\date{Submitted August 26, 2022; accepted October 30, 2022}

\abstract
{
SDSS-IV APOGEE-2, GALAH, and Gaia-ESO are high-resolution, ground-based, multi-object spectroscopic surveys providing fundamental stellar atmospheric parameters and multiple elemental abundance ratios for hundreds of thousands of stars of the Milky Way. Data from these and other surveys contribute to investigations of the history and evolution of the Galaxy.}
{We undertake a comparison between the most recent data releases of these surveys to investigate the accuracy and precision of derived parameters by placing the abundances on an absolute scale. We also discuss the correlations in parameter and abundance differences as a function of main parameters. Uncovering the variants provides a basis to continue the efforts of future sky surveys.}
{Quality samples from the APOGEE$-$GALAH (15,537 stars), APOGEE$-$GES (804 stars), and GALAH$-$GES (441 stars) overlapping catalogs were collected. We investigated the mean variants between the surveys, and linear trends were also investigated. We compared the slope of correlations and mean differences with the reported uncertainties.}
{The average and scatter of $v_{\rm rad}$, $T_{\rm eff}$, $\log g$, $\rm [M/H]$, and $v_{\rm micro}$, along with numerous species of elemental abundances in the combined catalogs, show that in general there is a good agreement between the surveys. 
We find large radial velocity scatters ranging from $1.3$~km/s to $4.4$~km/s when comparing the three surveys.
We observe some weak trends (e.g., in $\Delta T_{\rm eff}$ vs. $\Delta \log g$ for the APOGEE$-$GES stars)  and a clear correlation in the $v_{\rm micro}-\Delta v_{\rm micro}$ planes in the APOGEE$-$GALAH common sample.
For [$\alpha$/H], [Ti/H] (APOGEE$-$GALAH giants), and [Al/H] (APOGEE$-$GALAH dwarfs) potential strong correlations are discovered as a function of the differences in the main atmospheric parameters, and we also find weak trends for other elements.}
{In general we find good agreement between the three surveys within their respective uncertainties. However, there are certain regimes in which strong variants exist, which we discuss. There are still offsets larger than $0.1$~dex in the absolute abundance scales.}

\keywords{techniques: spectroscopic -- 
  techniques: radial velocities -- 
  Galaxy: abundances -- 
  Galaxy: evolution -- 
  Galaxy: fundamental parameters -- 
  galaxies: general}

\titlerunning{Comparison of APOGEE, GALAH, and GES}
\authorrunning{Heged\H{u}s et al. 2022}
\maketitle

\section{Introduction}\label{intro}

\begin{figure*}
\centering
\includegraphics[width=\textwidth]{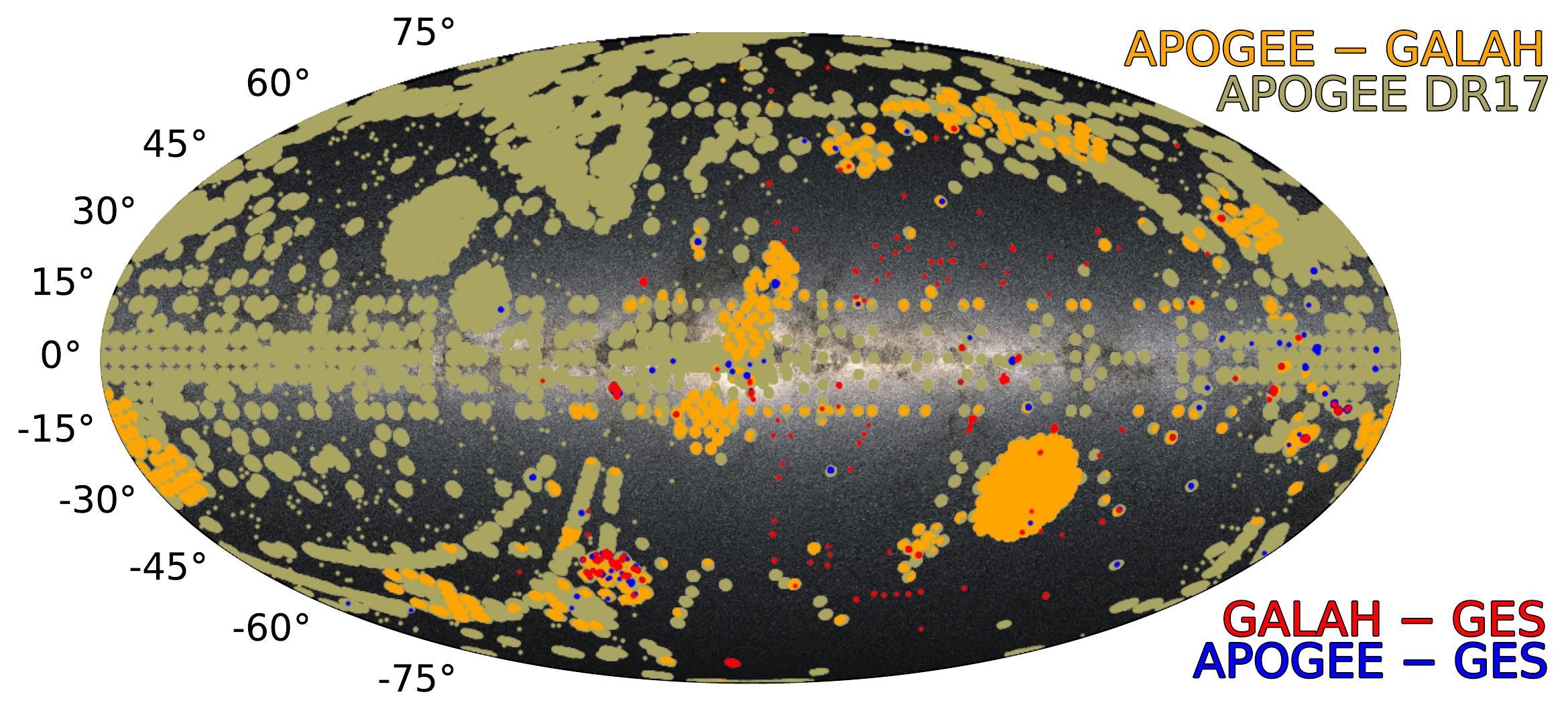} 
\caption{Mollweide projection of the Milky Way with the distribution of all targets from APOGEE DR17 (olive green) and from the  APOGEE$-$GALAH (orange), APOGEE$-$GES (blue), and GALAH$-$GES (red) common catalogs shown in Galactic coordinates.}
\label{gal_lb}
\end{figure*}

Starting in the 2010s several high-resolution stellar spectroscopic surveys started to map the chemical composition of our Galaxy in order to better understand its formation, structure, and evolution. Initially, the lower resolution surveys of the  Large sky Area Multi-Object fiber Spectroscopic Telescope \citep[LAMOST, ][]{cui_2012} and Radial Velocity Experiment \citep[RAVE, ][]{steinmetz_2020} focused mostly on the radial velocity of stars, while the later high-resolution multi-object programs such as the Gaia-ESO Survey \citep[GES, ][]{gilmore_2012, randich_2022}, Apache Point Observatory Galactic Evolution Experiment \citep[APOGEE-1 and APOGEE-2, ][]{majewski_2017}, and Galactic Archaeology with HERMES \citep[GALAH, ][]{desilva_2015} measured the chemical composition of hundreds of thousands of stars with higher precision and accuracy than the lower resolution surveys. Rigorous constraints of evolutionary models of the Milky Way generally require a combination of these abundances with the extremely precise astrometric and photometric measurements of the Gaia space mission \citep{gaia_2016,gaia_2018,gaia_2021}.
The Milky Way Mapper project \citep[MWM, ][]{kollmeier_2017} in the fifth iteration of the Sloan Digital Sky Survey (SDSS-V), William Herschel Telescope Enhanced Area Velocity Explorer \citep[WEAVE, ][]{dalton_2018}, Multi-Object Spectrograph Telescope \citep[4MOST, ][]{dejong_2019}, and Subaru Prime Focus Spectrograph \citep[PFS, ][]{takada_2014} are additional  upcoming surveys continuing to contribute to our understanding and theories of the Galactic chemo-dynamical evolution and history. As the successor of APOGEE, MWM will target more than five million stars across the Milky Way, collecting both near-infrared and optical spectra, with a significant portion of the total number of stars planned to be observed in both wavelength regimes. The stellar parameters and elemental abundances provided by this survey will enable a unique global Galactic map of the fossil records and chemo-dynamical structure. This will enable the quantitative testing of galaxy formation physics \citep{kollmeier_2017}. Here we  use the APOGEE DR17 results as the reference frame for future surveys, particularly the MWM currently underway.

The APOGEE-2 survey \citep{majewski_2017} is part of SDSS-IV \citep{blanton_2017}; its latest data release \citep[DR17, ][]{abdurro_2021}   has published the main atmospheric parameters and abundances of 20 species for 733,901 stars across the entire sky \citep{beaton_2021,santana_2021}. The APOGEE team has published both raw and calibrated effective temperatures and  surface gravities, and has applied small offsets to individual abundance ratios based on solar   neighborhood stars in DR17. The original APOGEE spectrograph (300 fibers) was placed on the 2.5-meter Sloan Foundation Telescope \citep{gunn_2006} at Apache Point Observatory (APO), and a clone APOGEE spectrograph was subsequently made and positioned (in early 2017) on the 2.5-meter Irénée du Pont Telescope \citep{bowen_1973} of Las Campanas Observatory (LCO) in Atacama de Chile \citep{wilson_2019}.  APOGEE is unique among the large spectroscopic surveys because it observes stars in the  near-infrared, specifically from 15140~\AA~to 16940~\AA\ with the resolving power of $R\!\!\sim$22,500, which allows it to reach the dust-obscured regions of the Milky Way in the disk, center, and bulge.

The GALAH observations are conducted with the multi-object 400-fiber HERMES spectrograph \citep{sheinis_2015} at the 3.9-metre Anglo-Australian Telescope in the Southern Hemisphere.  The content of its latest data release (DR3) is described by  \citet{buder_2021}. GALAH has published abundances for 30 species of elements in DR3. 
The GALAH survey obtains spectra in the following four optical windows: 4713$-$4903~\AA, 5648$-$5873~\AA, 6478$-$6737~\AA, and 7585$-$7887~\AA~ \citep{desilva_2015}. The resolving power is $\sim$28,000 in all four optical bands. GALAH DR3 contains detailed atmospheric parameters and abundances for 588,571 stars.

The Gaia-ESO Survey \citep{gilmore_2012,randich_2022} has made use of the capabilities of the FLAMES spectrographs \citep{pasquini_2002} mounted on the 8m ESO Very Large Telescope (VLT). This survey is unique in that it uses multiple pipelines (by various working groups) throughout the analysis of the same set of measurements, which is then  followed by data homogenization. GES has multiple narrow optical bands from 4033~\AA~ to 9001~\AA, and the resolution varies between $\sim$16,000 and 26,000 for GIRAFFE and is around $\sim$47,000 for the UVES spectrograph \citep{sacco_2014}. The latest GES catalog is DR5 \citep{randich_2022}; it contains stellar parameters and chemical data for 114,324 targets in the   southern sky. To date, GES   has made public the abundances of 31 elements.

Given the nature of these three different data sets, different methodologies were employed, leading to different systematic uncertainties associated with each of the analyses. Comparing the results in common between these data sets helps us to understand systematics and correlations, and potentially correct them for future programs. Based on metallicity measurements from open clusters, \citet{spina_2022} performed the homogenisation of the  APOGEE, GALAH, and GES high-quality data sets 
which was analyzed as a whole, for example to trace the radial metallicity distributions in the Galaxy.

In this paper we therefore aim to determine the differences and underlying systematic effects between the latest data releases of three independent high-resolution spectroscopic surveys: APOGEE DR17, GALAH DR3, and GES DR5. As it is widely understood, abundances on the absolute scale not only depend on the selected line lists and continuum placement, but also strongly  on the theoretical assumptions made in the construction of the model atmosphere and synthesis calculations  \citep[e.g.,][]{jofre_2019}. By combining tens of thousands of stars all observed by these surveys and with the application of several quality criteria, we can investigate the accuracy of the stellar atmospheric parameters and abundances of several elements on the absolute scale in a large effective temperature, surface gravity, and metallicity range covering the majority of the Hertzsprung--Russell Diagram (HRD). Understanding how the abundances differ between the various analyses feeds back into fine-tuning Galactic formation and evolutionary models.

This paper is structured as follows. In Sect.~\ref{target_data} we provide an overview of the  target and data selection, covering the determination of our parameter space, and the three resulting common quality data sets. Section~\ref{analysis} focuses on the methods and the steps we implemented in our analysis.
Radial velocity and stellar parameter differences, along with the application of all global filters, are explained in Sect.~\ref{rad_vel} and Sect.~\ref{disc_diff}, respectively. Individual elemental and $\alpha$-abundance ratios from the APOGEE$-$GALAH overlapping catalog are thoroughly discussed in Sect.~\ref{disc_diff_abund}. We briefly examine how these systematic differences translate to the existing chemical maps of our Galaxy in Sect.~\ref{mw_maps}. We go into detail about the effect of the performed APOGEE calibrations on the discovered variants in Sect.~\ref{calib}. Our conclusions are presented in Sect.~\ref{conclusion}.

\section{Target and data selection}\label{target_data}
\subsection{Target selection}\label{target}

\begin{figure}
\centering
\includegraphics[width=.4\textwidth,angle=0]{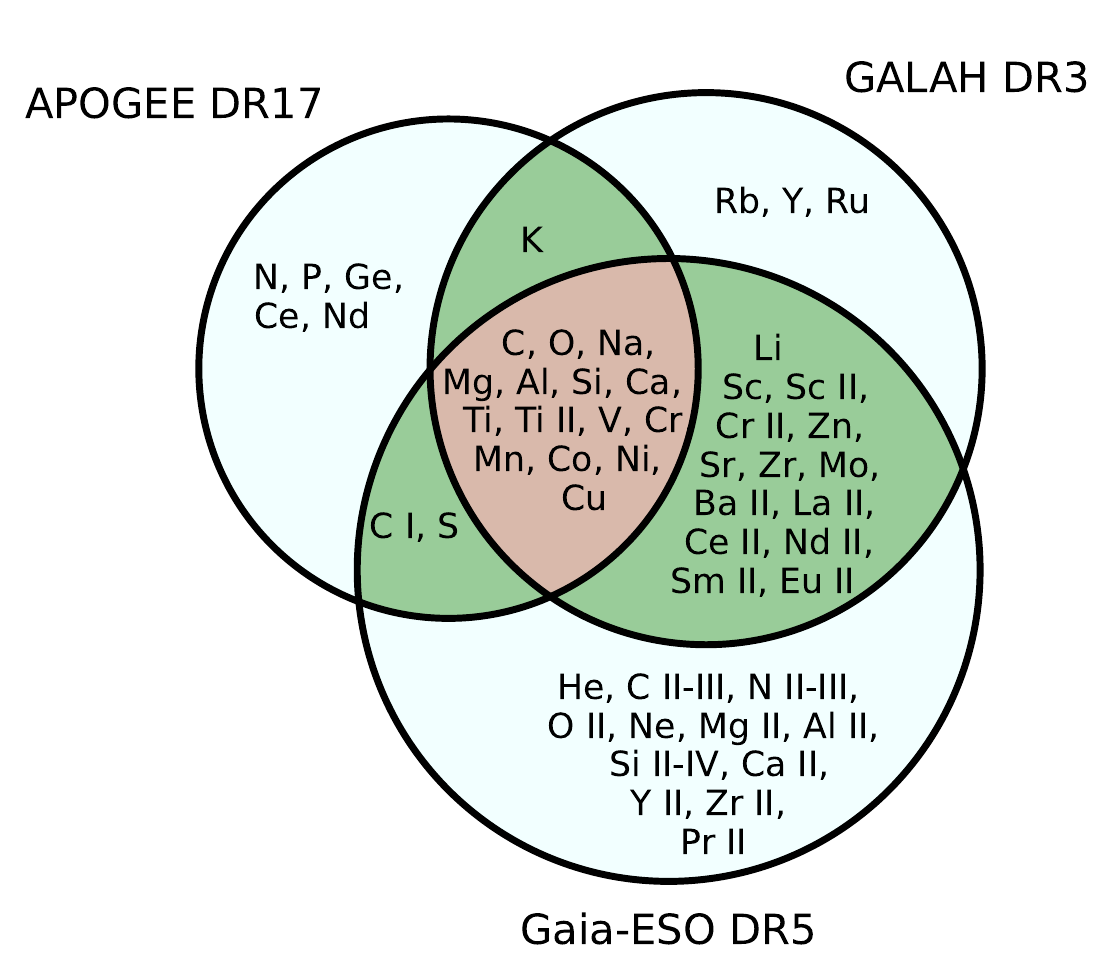}
\caption{Venn diagram of the elements derived by each survey. The overlap of elements in all three surveys is in brown and those from only two surveys in dark green. Light green indicates no overlap between surveys. Species of any intersection are discussed in detail in Sect.~\ref{disc_diff_abund}.}
\label{venn}
\end{figure}

\begin{scriptsize}
\begin{table*}
\caption{Details of the latest data releases of the  APOGEE, GALAH, and Gaia$-$ESO sky survey programs.}
\begin{tabular}{p{3cm}p{4.5cm}p{4.5cm}p{4.5cm}}
\hline\hline & 
{\textbf{APOGEE}} & 
{\textbf{GALAH}} &
{\textbf{GES}} \\
 & 
{DR17} &
{DR3} &
{DR5} 
\\\hline
Date of data release 
& December 2021 
& November 2020 
& May 2017
\\ 
Number of stars 
& 733,901 
& 588,571 
& 114,324 
\\ 
Telescope 
& Sloan Foundation at APO, 2.5~m NMSU at APO, 1~m
&  Anglo-Australian
Telescope at Siding Spring Observatory, 3.9~m 
&   ESO Telescopes at the La Silla Paranal Observatory
\\ 
& du Pont telescope at LCO, 2.5~m 
& & \\
Wavelength range 
& 15 140$-$16 940~\AA 
&  4 optical bands:   
& multiple ranges 
\\ 
& & 4713$-$4903~\AA, 5648$-$5873~\AA
& from 4033~\AA ~to 9001~\AA   \\
& &  6478$-$6737~\AA, 7585$-$7887~\AA
& \\
Resolution
& $\sim$ 22 500 
& $\sim$ 28 000 
&  $\sim$ 16~000 $-$ 26~000, 47~000 \\ 
List of parameters 
& $T_{\rm eff}$, $\log g$, $v_{\rm rad}$, [M/H], [$\alpha$/Fe], [Fe/H], $ v_{\rm micro}$, $v\sin{i}$(giants: $v_{\rm mac}$)
& $T_{\rm eff}$, $\log g$, $v_{\rm rad}$, [Fe/H], [$\alpha$/Fe], $v_{\rm micro}$, $v_{\rm broad}$ 
& $T_{\rm eff}$, $\log g$, $v_{\rm rad}$, [Fe/H], $v_{\rm micro}$, $v\sin i$
\\
List of elements 
& C I, N, O, Na, Mg, Al, Si, P, S, K,
Ca, Ti, Ti II, V, Cr, Mn, Co, Ni, Cu, Ge, Ce,
Nd 
& Li, C, O, Na, Mg, Al, Si, K, Ca, Sc~I, 
Ti, Ti II, V, Cr, Cr II, Mn, Co, Ni, Cu, Zn, Rb, Sr, Y, Zr, Mo, Ru,   Ba II, La II, Ce II, Nd~II, Sm~II, Eu II
& He, Li, C~I-III,  N~II-III,  O~I-II, Ne, Na, Mg, Al~I-II, Si~I-IV, S, Ca~I-II, Sc~I-II, Ti, Ti~II, V, Cr~I-II, Mn, Co, Ni, Cu, Zn, Sr, Y~II, Zr~I-II, Mo,  Ba II, La II, Ce II, Pr~II, Nd~II, Sm~II, Eu II
\\
Calibration 
& calibrated $T_{\rm eff}$, $\log g$, [M/H], [$\alpha$/M], abundance parameters
& calibrated abundance parameters
& \textit{Gaia} benchmark stars and calibrating clusters
\\
Method of deriving parameters 
& ASPCAP (parameters and abundances)
& Spectrum synthesis (SME and \textit{The Cannon})
& equivalent widths, comparisons of observed spectra and a grid of templates (strategy: WGs) \\
Model atmospheres 
& Kurucz model (Cannon models), all-MARCS grid ($T_{\rm eff}<$8000~K)
& MARCS grid (plane parallel models for $\log g>4$, spherical models for $\log g <4$)
& MARCS grid
\\
1D/3D 
& spherical (Turbospectrum) or plane parallel (Synspec)
& 1D spherical
& 1D spherical \\
LTE/NLTE 
& LTE/NLTE 
& LTE/NLTE & \\
Spectral synthesis code 
& Turbospectrum (for the main spectral grids), Synspec
& Spectroscopy Made Easy (SME)
& e.g. Turbospectrum, SME, MOOG, COGs
\\
1D/3D 
& Plane parallel and spherical
& 1D spherical
& \\
LTE/NLTE 
& LTE/NLTE 
& LTE/NLTE 
& LTE \\
Solar reference table 
&  \citet{grevesse_2007} 
& \citet{buder_2021}
& {\citet{grevesse_2007}} \\
\hline
\label{sum_apogalges}
\end{tabular}
\end{table*}
\end{scriptsize}

In this section we describe how the SDSS-IV APOGEE-2 DR17 \citep{abdurro_2021}, GALAH DR3 \citep{buder_2021}, and Gaia-ESO DR5 \citep[iDR6, ][]{randich_2022} data sets were combined   and how the overlapping stellar catalogs were created. We also briefly discuss the features of these common data sets, as well as the atmospheric parameters derived by these sky surveys. 

The main stellar parameters derived by all three surveys  investigated in this paper are radial velocity ($v_{\rm rad}$), effective temperature ($T_{\rm eff}$), surface gravity ($\log g$), overall metallicity ([M/H]$=$[Fe/H]), and microturbulent velocity ($v_{\rm micro}$). In addition to 
these common parameters, APOGEE published [$\alpha$/Fe] and $v\sin i$ ($v_{\rm mac}$ for giants), GALAH derived [$\alpha$/Fe] and $v_{\rm broad}$, and GES calculated $v\sin i$. The full list of individual parameters and those present in the overlapping data sets are listed in Tables~\ref{sum_apogalges} and \ref{sum_apogalges2}, respectively.
As listed in the relevant column of \autoref{sum_apogalges}, 20 individual elemental abundances (including C and C~I, Ti~I and Ti~II) were released as part of DR17 by APOGEE. 
From the latest data release of GALAH (DR3), the 30 species of abundances are listed in the second column of \autoref{sum_apogalges}.
In its fifth data release, GES has made public abundances of 31 elements, which are listed in the third column of \autoref{sum_apogalges}. The description of this data release can be found in an ESO document\footnote{ESO Phase 3 Data Release 5 Description: \url{https://www.eso.org/rm/api/v1/public/releaseDescriptions/191}} and in \citet{randich_2022}.

In order to create the APOGEE$-$GALAH, APOGEE$-$GES, and GALAH$-$GES stellar catalogs, we combined the published data from APOGEE DR17, GALAH DR3, and GES DR5 using \texttt{TOPCAT} version 4.8-2 (Tool for OPerations on Catalogues And Tables; \citet{topcat_2005}). 
Before matching stars, we removed all stars with \texttt{STAR\_BAD} flags from the APOGEE DR17 sample. We performed a 1$''$ cross-match in \texttt{TOPCAT} and then removed GALAH and GES stars with flags indicating one or more problematic parameters. More details about the flagging systems and our additional filtering aspects are given in Sect.~\ref{data}. Entries with APOGEE \texttt{STAR\_BAD} flags were also removed (see Sect.~\ref{data} for  discussion).

The combined APOGEE$-$GALAH data set contains 37,770 stars, the APOGEE$-$GES catalog 2,502 stars, and the GALAH$-$GES common sample 1,510 stars before applying other quality cuts to the data.
{We define these high-quality data sets as golden samples. }
The positions in Galactic coordinates $(l,b)$ of all APOGEE stars and the positions from the three common data sets are shown in \autoref{gal_lb}. While APOGEE often publishes multiple values of the parameters, if the particular star was part of multiple science programs, none of the stars in our common sample had multiple entries in the final DR17 catalog. 

The stars observed in APOGEE are from both hemispheres, and these targets are  located in the highly reddened Galactic disk and bulge, and in the Galactic halo \citep{zasowski_2017,beaton_2021,santana_2021}. Interstellar extinction is less significant in the near-infrared H-band than in the optical (A(H)/A(V)$\sim$1/6, \citet{cardelli_1989}), which allowed APOGEE to probe the Galactic thin and thick disks in more detail than GALAH and GES. 
In addition to  observing the Milky Way, the  APOGEE program also included the Magellanic Clouds (MCs), eight dwarf spheroidal satellites, and  observations of both M33 and M31 \citep{beaton_2021,santana_2021}.
GALAH observes stars from the southern hemisphere, mostly along the ecliptic or Galactic latitudes lower than $\pm$30$^\circ$ and the MCs, but not in the thin disk, avoiding high-extinction areas. Stars in GES are from the southern hemisphere along the ecliptic, and field stars in the Galactic halo, the thick and thin disks, and the Galactic bulge. 

While the three surveys computed chemical abundances for many elements, not all elements can be studied by each survey. The common parameters and species between APOGEE, GALAH, and GES
can be found in \autoref{sum_apogalges2}. Between APOGEE and GALAH the common derived elemental abundances are C, O, Na, Mg, Al, Si, K, Ca, Ti, Ti~II, V, Cr, Mn, Co, Ni, and Cu. Between APOGEE and GES we are able to compare the   species  C, O, Na, Mg, Al, Si, S, Ca, Ti, Ti~II, V, Cr, Mn, Co, Ni, and Cu. From the GALAH-GES sample we can examine differences in the abundances of Li, C, O, Na, Mg,  Al, Si,  Ca, Sc, Sc~II, Ti, Ti~II, V,  Cr, Cr~II, Mn, Co, Ni, Cu, Zn, Sr, Zr, Mo, Ba~II, La~II, Ce~II, Nd~II, Sm~II, and Eu~II. 
\autoref{venn} shows a diagram of all elements appearing in at least one survey's parameter list.

\subsection{Selecting the parameter space}\label{data}

In this section we describe the quality cuts applied to each survey 
data set to ensure that unreliable measurements are not included in the comparisons, so that we can reliably explore the underlying systematic differences.

We applied several global and survey specific local cuts for each survey, which are defined here. 
Global cuts are applied on the samples before any calculation, and affect all parameters. {Moreover,} we use local cuts and flags where individual abundances are considered, which  makes the analysis of discrepancies of each elemental abundance independent.

In the first step we  filtered out APOGEE stars that have the quality control flag \texttt{STAR\_BAD} set, but retained stars with the \texttt{STAR\_WARN} flag. This  corresponded to the  bit 23 part of \texttt{APOGEE\_ASPCAPFLAG} bitmask. The  \texttt{STAR\_BAD} flag is set for a star if any of \texttt{TEFF\_BAD}, \texttt{LOGG\_BAD}, \texttt{CHI2\_BAD}, \texttt{COLORTE\_BAD}, \texttt{ROTATION\_BAD}, \texttt{SN\_BAD} or \texttt{GRIDEDGE\_BAD} occur\footnote{\url{https://www.sdss.org/dr17/irspec/apogee-bitmasks/} \\The web documentation contains the details of the APOGEE Bitmasks. This documentation also includes detailed flag descriptions.} \citep{abdurro_2021}.
We also selected APOGEE stars within the following global criteria: signal-to-noise ratio S/N$>$100 
per half-resolution element per pixel to  minimize uncertainties caused by random noise; \texttt{VSCATTER}$<$1~km/s for the scatter in the radial velocity (RV) to eliminate most binaries and variable stars;  \texttt{VERR}$<$1~km/s for the error of RV to discard potentially bad measurements; and \texttt{VSINI}$<$15~km/s for the rotational velocity to avoid significant rotational broadening.

From the GALAH survey we selected stars that satisfy the following criteria: S/N$>$30 (\texttt{snr\_c3\_iraf}$>$30, as recommended in GALAH DR3 documentation\footnote{\url{https://www.galah-survey.org/dr3/flags/}}); \texttt{VBROAD}$<$15~km/s to discard significant stellar rotation. In addition, only stars with quality flags   \texttt{flag\_sp}=0 (global), \texttt{flag\_fe\_h}=0 (global), and \texttt{flag\_X\_fe}=0 (local) were selected to avoid identified problems with stellar parameter determination, metallicity, and abundance determination of element [X/Fe], respectively (as advised in GALAH DR3 documentation, and  discussed in \citealt{buder_2021}).

The GES data set was filtered by utilizing the survey's flagging system and recommendations defined in ESO Phase 3 Data Release 5 Description, Tables 3-7. We list the criteria, and we report the values of the eliminated flags of \texttt{SLFAGS} column in brackets. We removed stars with a relative error in effective temperature larger than 5\%, with  $\texttt{S/N}<50$ (and \texttt{SFLAG==SNR}), with RV error of $\texttt{E\_VRAD}>1$~km/s and with significant rotation: $\texttt{VSINI}>15$~km/s  (and \texttt{SFLAG==SRO}). We also removed stars with unreliable RV results that are presumable binary or variable stars (\texttt{SRV, BIN}), stars with spectral reduction issues (\texttt{SRP}), stars that are out of grid boundaries (\texttt{PSC}), and  spectra containing any emission lines (e.g., H$\alpha$)  (\texttt{EML}).

In addition to  the considerations listed above, stars with missing parameter values in their respective databases were removed. When analyzing specific individual abundances stars are excluded if the particular abundance value of that element is not reported or flagged.

\subsection{Common data description} \label{common_descr}

\begin{figure*}
\centering
\includegraphics[width=\textwidth,angle=0]{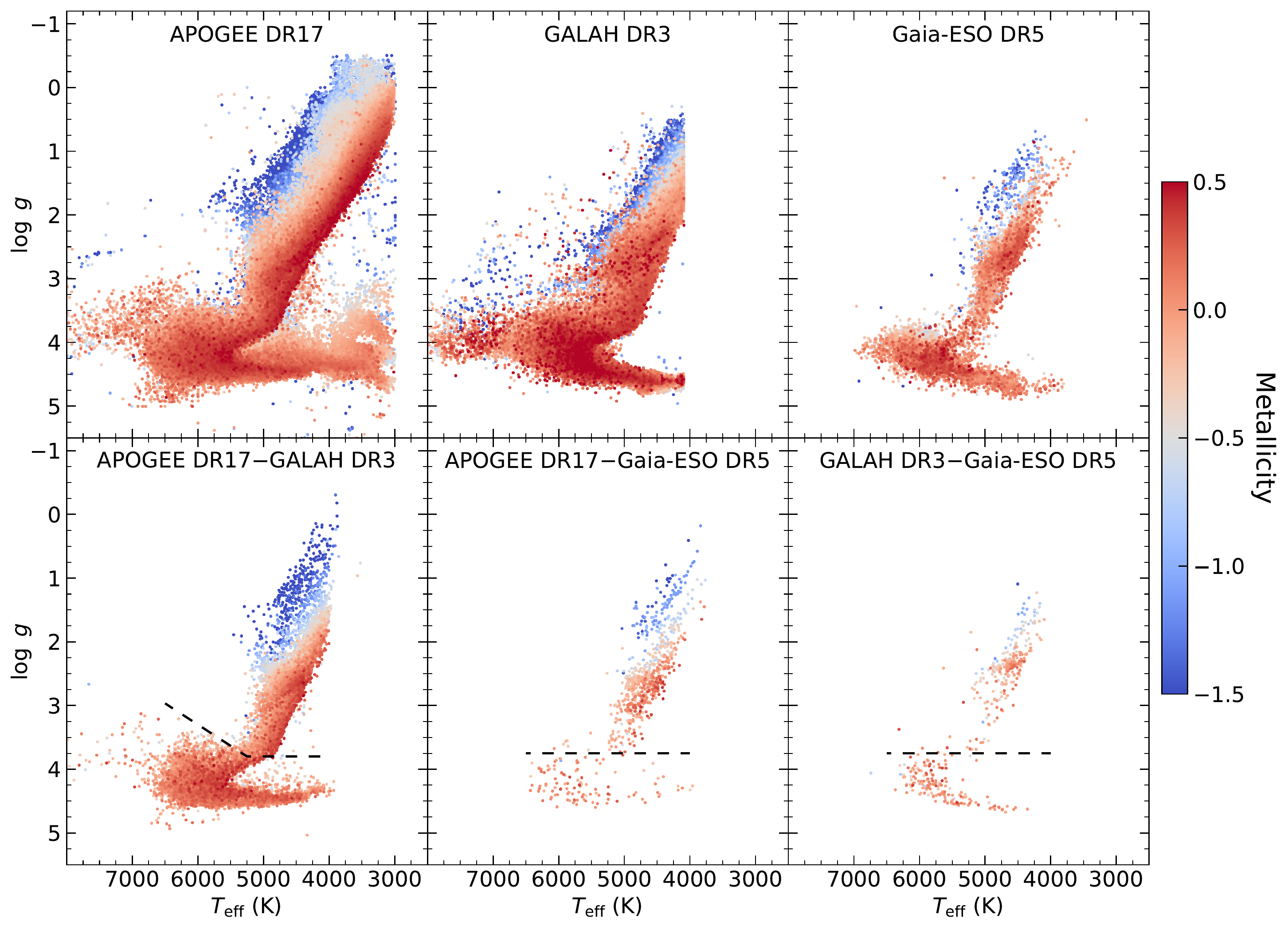}
\caption{$T_{\rm eff}-\log g$ diagrams of quality stars observed in APOGEE DR17, GALAH DR3, GES DR5 surveys (top row) and of stars that are common between these surveys (bottom row) color-coded by metallicity. 
In the APOGEE$-$GALAH and APOGEE$-$GES panels in the bottom row the APOGEE DR17 $T_{\rm eff}$ (K), $\log g$ (cgs, dex), and [M/H] (dex) are used on the axes and in the color-coding;  in the  GALAH$-$GES panel the GALAH DR3 $T_{\rm eff}$, $\log g$, and [Fe/H] are used. The separations of dwarf and giant stars are indicated by dashed lines in the common data sets.}
\label{teff_logg}
\end{figure*}

Before discussing the discrepancies in atmospheric parameters and abundance ratios, we briefly describe the characteristics of the selected common data sets obtained after cuts. 

Quality cuts left 15,537 stars, 804 stars, and 441 stars in the APOGEE$-$GALAH, APOGEE$-$GES, and GALAH$-$GES common catalogs, respectively. 
\autoref{teff_logg} shows $T_{\rm eff}$ versus $\log g$ (Kiel) diagrams. The top row  presents the full set of stars from each individual survey separately. The bottom row  contains the diagrams of common stars from the APOGEE$-$GALAH, APOGEE$-$GES, and GALAH$-$GES catalogs color-coded by metallicity values after all cuts have been applied.
GALAH stars with $T_{\rm eff} < 4100$~K were filtered out by the flagging system implemented by GALAH. Using Kiel diagrams it is relatively easy to separate the main sequence  (MS) and red-giant branch (RGB) stars. We defined APOGEE$-$GALAH stars to be giants using the following equations: $T_{\rm eff} < 5250 \rm ~K$ and $\log g < 3.80$~dex, or $T_{\rm eff} > 5250 \rm ~K$ and $\log g + (T_{\rm eff}-4950~K)/1500 ~{\rm K} < 4$~dex; in the other two cases giants are those with $\log g < 3.75$~dex irrespective of $T_{\rm eff}$, as indicated by the dashed lines in \autoref{teff_logg} plots in the bottom row of plots. The reason behind having different definitions for the three overlapping data sets is that they are different data sets and have their own small offsets, so a cut that is suitable for one data set may not work for another in practice. After applying global cuts the APOGEE$-$GALAH common stellar catalog contains 7,763 dwarfs and 7,774 giants; there are 112 dwarfs and 692 giants in APOGEE$-$GES; and there are 170 dwarfs and 271 giants  in GALAH$-$GES. The common data description outlined in this section generally covers all the analyses of main atmospheric parameters (see Sect.~\ref{disc_diff}). 

It should be noted that the APOGEE$-$GALAH sample covers a large range of 4000~K$-$7000~K and 0$-$4.75~dex in $T_{\rm eff}$ and  $\log~g$, respectively. As a result of the smaller samples they contain, the APOGEE$-$GES and GALAH$-$GES common samples seem to have a narrower distribution in the parameter space. Moreover, in the last two cases the number of MS stars is relatively low.

\section{Methods of surveys}\label{analysis}

Throughout its data releases, the APOGEE survey provided both ``raw'' parameters, derived from the  APOGEE Stellar Parameters and Chemical Abundance Pipeline  \citep[ASPCAP;][]{garcia_2016}, which relies on the   FERRE\footnote{\url{github.com/callendeprieto/ferre}} multidimensional $\chi^2$ optimization code \citep{allende_2006}, and from the APOGEE line list (Smith et al. 2021), and also parameters  calibrated to values that were independently derived. 
Because we are interested in how well the various methods used to derive parameters and abundances behave, we decided to use the APOGEE raw parameters and abundance values for the discussions presented in this paper. A brief description of the  calibrations is provided in  Sect.~\ref{calib} where the effect of the APOGEE $T_{\rm eff}$, $\log g$, and [M/H] calibrations on the discrepancies between the surveys is studied specifically. 

The raw values of $T_{\rm eff}$, $\log g$, [M/H], and $v_{\rm micro}$ used here were taken from the \texttt{FPARAM} array, while the uncalibrated individual abundances are listed under the \texttt{FELEM} label \citep{abdurro_2021}. We note that elemental abundances are determined relative either to hydrogen ([X/H]) or iron ([X/M]), and the conversion between them is given by the following relation:
\begin{equation}
    \rm [X/H] = [X/M] + [M/H].
\end{equation}
We note that APOGEE calculated both overall metallicity denoted by [M/H] and iron abundance [Fe/H] \citep{abdurro_2021}, while the other two surveys use the iron abundance as a proxy for metallicity. 
The two metallicity indicators generally provide the same values within the expected uncertainties. Here, a mean difference between iron and all metal abundances of $0.002\pm0.011$~dex is obtained for the APOGEE$-$GALAH catalog. We discuss further details about the differences between the various definitions 
in Sect.~\ref{disc_met}. 

In terms of the output of GALAH DR3, the elemental abundances are derived relative to Fe, while in GES DR5 the provided values are absolute abundances. Generally in our calculations, we also apply the 
\begin{equation}\label{abs_abund}
    \rm A(X) = [X/H] + A(X_\odot) 
\end{equation}
relation, where [X/H] is the relative abundance and A(X) is the absolute abundance, for which the related solar reference value is A(X$_\odot$). 

The absolute abundance is a definition based on how the solar reference table is constructed, and these values vary from survey to survey, as indicated in the last row in \autoref{sum_apogalges}. Our main motivation is to explore the abundance differences between the results for the three surveys on an absolute scale.  {APOGEE and GES reported the use of \citet{grevesse_2007} as solar reference for the abundances.} In addition, GALAH derived their own solar reference abundances; these zero points are published in \citet{buder_2021}, and we note that there are several elements with multiple reference values reported for each absorption line used in the fitting instead of a combined one. We  therefore   use the mean of these values in those cases.

\section{Comparison of radial velocities}\label{rad_vel}

The radial velocities (RV) in the three surveys are calculated first with the help of preliminary estimated effective temperatures and surface gravities before deriving the final parameters for each star. In order to assess the magnitude of the differences and scatter between surveys, we compare them to the average uncertainties reported by each survey.

In APOGEE DR17 the data reduction algorithm \citep{nidever_2015} performs a least-squares fit to a set of visit spectra, solving simultaneously for basic stellar atmospheric parameters and the radial velocity for each visit. The measurements are reported under the  \texttt{VHELIO\_AVG} label and corrected for the barycentric motion \citep{abdurro_2021}. The overall average of RV uncertainty, $\sigma_{v_{\rm rad}}$ for APOGEE DR17 quality stars is reported in \autoref{err} (both for dwarfs and giants shown in the top left panel of \autoref{teff_logg}).

For each GALAH DR3 spectrum, RV is fitted as part of the spectrum synthesis comparison and has the barycentric correction applied. For analyses we use preferably \texttt{rv\_nogr\_obst} first, and if there is no entry for that parameter we use \texttt{rv\_sme\_v2}, which are not corrected for gravitational redshifts \citep{buder_2021}. 
The average uncertainties of individual RVs in our quality GALAH sample is provided in \autoref{err}. 

In GES DR5 RVs are derived by cross-correlating each spectrum with a grid of synthetic template spectra. In the quality GES DR5 data set the average uncertainty of RVs is listed in \autoref{err}.

\begin{figure*}
\centering
\includegraphics[width=\textwidth,angle=0]{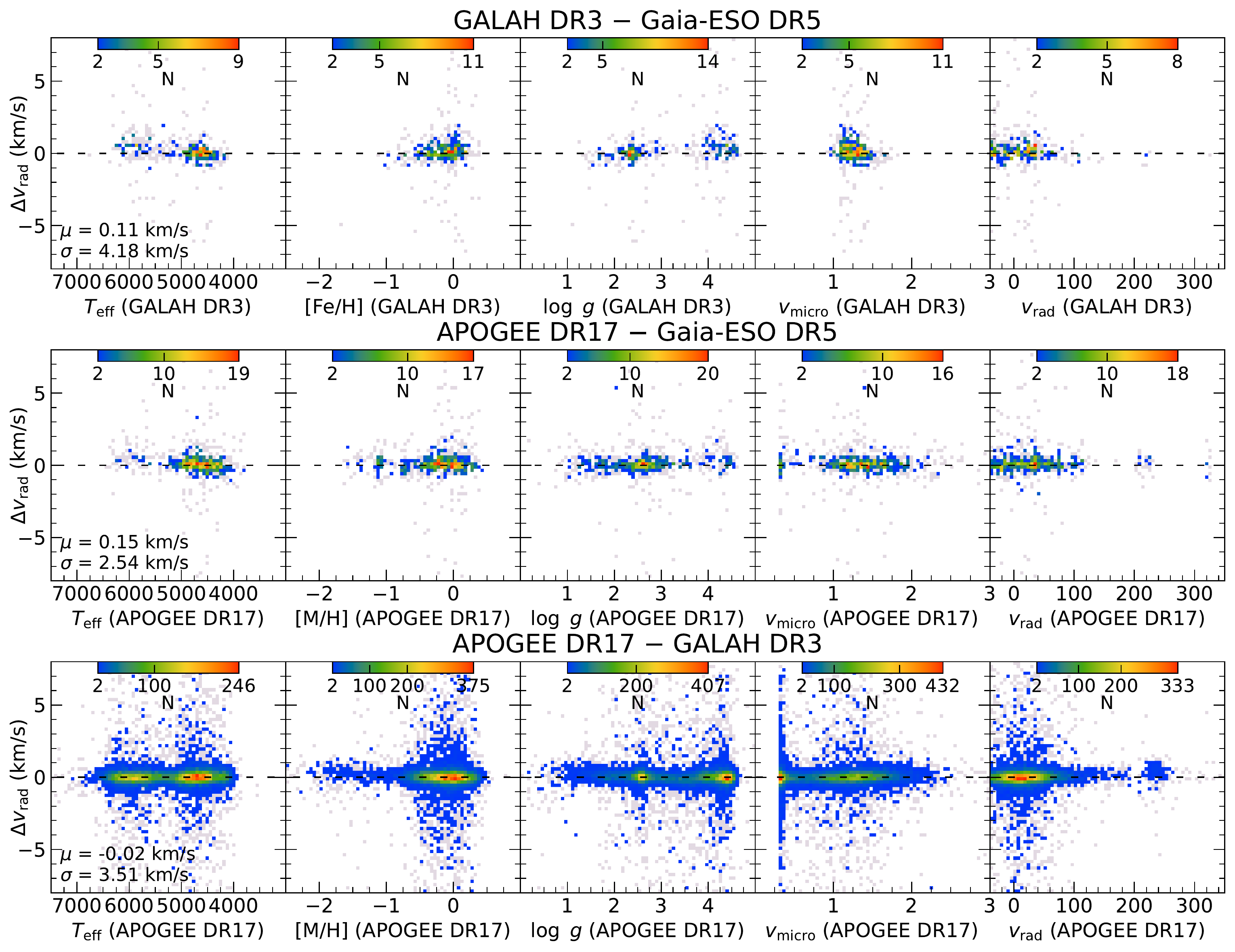}
\caption{Distributions of the radial velocity differences with respect to the main stellar parameters for the GALAH$-$GES (top row), APOGEE$-$GES (middle row), and APOGEE$-$GALAH (bottom row) common data sets, including dwarfs and giants.  Bins containing only one star are in gray  in the background; the color-coding (according to the number of stars per bin) starts from  two stars. The mean and standard deviation of the differences are also shown in each case. Horizontal axes indicate $T_{\rm eff}$ (K), metallicity, $\log g$ (dex), and  $v_{\rm micro}$ and $v_{\rm rad}$ (km/s) from APOGEE DR17 (middle and bottom rows) and from GALAH DR3 (top row).}
\label{param_drv}
\end{figure*}

It should be noted that the RV difference, $\Delta v_{\rm rad}$, is calculated as $v_{\rm rad(APOGEE)} - v_{\rm rad(GALAH)}$, $v_{\rm rad(APOGEE)} - v_{\rm rad(GES)}$, and $v_{\rm rad(GALAH)} - v_{\rm rad(GES)}$ for the APOGEE$-$GALAH, APOGEE$-$GES, and GALAH$-$GES catalogs, respectively. 
Radial velocity differences as functions of $T_{\rm eff}$, metallicity, $\log g$, $v_{\rm micro}$, and $v_{\rm rad}$ are shown in \autoref{param_drv}. 
We do not distinguish MS stars from red giants in \autoref{param_drv}, but \autoref{sum_apogalges2} also contains calculations of the average and scatter of RV discrepancies for dwarfs and giants separately. For all stars in common after quality cuts between APOGEE and GALAH, a $v_{\rm rad}$ discrepancy of $-0.02\pm3.51$~km/s is found (see \autoref{param_drv} bottom row), and $95.0\%$ of these stars are located in its $\pm1\sigma$ confidence range. We find no clear significant correlations in RV measurements as a function of atmospheric parameters, except for a weak trend for metal-poor red giants (bottom row, second panel from the right for [M/H]$<-1.0$, \autoref{param_drv}). When separating MS dwarfs from RGB  giants, the offset is less significant for dwarfs ($-0.01\pm4.10$~km/s) than for giants ($-0.04\pm2.81$~km/s). These offsets do not indicate systematic errors as they are significantly smaller than the estimated dispersion of both APOGEE and GALAH. Although the discrepancy distributions are not exactly Gaussian,  the $\sigma_{v_{\rm rad}}$ uncertainties are combined by using the square root of the sum of squares. Therefore, for the APOGEE$-$GALAH golden sample the estimated RV internal uncertainty is $0.12$~km/s.

Although the global cuts have already been applied, we note that the number of stars with an absolute RV difference larger than $1$~km/s is $1608$. The most likely candidates would be variable stars (pulsators, binaries, eclipsing systems) that remained in the filtered quality sample. In order to clarify this idea, we matched the APOGEE$-$GALAH table with the APOGEE DR17 Value Added Catalog (VAC) of double-lined spectroscopic binaries \citep{kounkel_2021}, and with the General Catalog of Variable Stars \citep[GCVS, ][]{samus_2017}. After cross-matching our sample with these catalogs, only six matches were found in the two cases   together. Global cuts included the removal of stars with a GALAH \texttt{flag\_sp} that indicates stars with likely unreliable astrometry in Gaia, defined as $\texttt{ruwe\_dr2}>1.4$. Below this value of the  RUWE parameter we find no significant correlation between $\Delta v_{\rm rad}$ and \texttt{ruwe\_dr2}.

Furthermore, we made use of the Variable Summary of Gaia DR3 \citep{eyer_2022}. A \texttt{source\_id}-based  cross-match between the APOGEE$-$GALAH quality data set and this catalog resulted in $437$ common stars. The combined RV uncertainty of APOGEE and GALAH is $0.12$~km/s (calculated with the values from \autoref{err}), and we conclude that only $2.24$\% of the matched stars lie outside of this uncertainty range. 
Therefore, the reason for this variation is unclear, while we can conclude that our global cuts that remove known variables from the samples have been effective. The presence of as-yet-unknown measurement discrepancies or unidentified variables might explain this issue.
In addition, \citet{anguiano_2018} have found similarly large differences in RVs from comparing APOGEE DR14 and LAMOST DR3.

\begin{scriptsize}
\begin{table*}
\begin{threeparttable}
\caption{Comparison of the common catalogs of the surveys. The quality cuts, common parameters, and abundances are listed for the  APOGEE$-$GALAH, APOGEE$-$Gaia-ESO, and GALAH$-$Gaia-ESO stellar catalogs. The average and scatter of differences are calculated overall, and  after separating MS and GB stars. No data is given in the table cell if there is no available corresponding data. The confidence interval is $\pm 1 \sigma$. Overall means do not differentiate between MS and RGB stars.}
\begin{tabular}{lrrrrrr}
 \hline\hline & 
\multicolumn{2}{c}{\textbf{APOGEE-GALAH}} & 
\multicolumn{2}{c}{\textbf{APOGEE-GES}} &
\multicolumn{2}{c}{\textbf{GALAH-GES}} \\
\hline
\multicolumn{1}{p{2.0cm}}{No. common stars\tnote{a}}
& \multicolumn{2}{p{3.8cm}}{37,770 (15,537)} 
& \multicolumn{2}{p{3.8cm}}{2,502 (804)}  
& \multicolumn{2}{p{3.8cm}}{1,510 (441)} \\ 
\multicolumn{1}{p{2.0cm}}{Common parameters} 
&  \multicolumn{2}{p{3.8cm}}{$T_{\rm eff}$, $\log g$, $v_{\rm rad}$, [Fe/H], [$\alpha$/Fe], $v_{\rm micro}$} 
&  \multicolumn{2}{p{3.8cm}}{$T_{\rm eff}$, $\log g$, $v_{\rm rad}$, [Fe/H], $v_{\rm micro}$}
& \multicolumn{2}{p{3.8cm}}{$T_{\rm eff}$, $\log g$, $v_{\rm rad}$, [Fe/H], $v_{\rm micro}$}  \\ 
\multicolumn{1}{p{2.0cm}}{Common elements} 
&  \multicolumn{2}{p{3.8cm}}{C, O, Na, Mg, Al, Si,   K, Ca, Ti, Ti~II, V, Cr, Mn, Co, Ni, Cu}  
&  \multicolumn{2}{p{3.8cm}}{C, C~I, O, Na, Mg, Al, Si, S, Ca, Ti, Ti~II, V, Cr, Mn, Co, Ni, Cu}
&   \multicolumn{2}{p{3.8cm}}{Li, C, O, Na, Mg,  Al, Si,  Ca, Sc, Sc~II, Ti, Ti~II, V,  Cr, Cr~II, Mn, Co, Ni, Cu, Zn, Sr, Zr, Mo, Ba~II, La~II, Ce~II, Nd~II, Sm~II, Eu~II}
\\ 
\multicolumn{1}{p{2.0cm}}{S/N cut\tnote{b}} & \multicolumn{2}{p{3.8cm}}{\texttt{SNR}$>$100 \& \texttt{snr\_c3\_iraf}$>$30} &  \multicolumn{2}{p{3.8cm}}{\texttt{SNR}$>$100 \& $\texttt{SNR}\!\!>\!\!50$} &  \multicolumn{2}{p{3.8cm}}{\texttt{snr\_c3\_iraf}$>$30 \& $\texttt{SNR}\!\!>\!\!50$} \\ 
\multicolumn{1}{p{2.0cm}}{Other quality cuts\tnote{b}} &  \multicolumn{2}{p{3.8cm}}{$\texttt{vbroad}\!\!<\!\!15$ \& $\texttt{flag\_sp}\!\!==\!\!0$ \& $\texttt{flag\_fe\_h}\!\!==\!\!0$ \& $\texttt{VSCATTER}\!\!<\!\!1$ \& $\texttt{VSINI}\!\!<\!\!15$ \& $\texttt{VERR}\!\!<\!\!1$, flags }& \multicolumn{2}{p{3.8cm}}{$\texttt{VSCATTER}\!\!<\!\!1$ \& $\texttt{VSINI}\!\!<\!\!15$ \& $\texttt{VERR}\!\!<\!\!1$ \& $\texttt{ERROR\_TEFF/TEFF}\!\!<\!\!5\%$ \& $\texttt{E\_VRAD}\!\!<\!\!1$, flags}   & \multicolumn{2}{p{3.8cm}}{$\texttt{ERROR\_TEFF/TEFF}\!\!<\!\!5\%$ \&  $\texttt{flag\_sp}\!\!==\!\!0$ \& $\texttt{flag\_fe\_h}\!\!==\!\!0$ \& $\texttt{vbroad}\!\!<\!\!15$ \& $\texttt{VSINI}\!\!<\!\!15$ \& $\texttt{E\_VRAD}\!\!<\!\!1$, flags}   \\ 
\hline
\hline
\multicolumn{1}{p{2.0cm}}{} & \multicolumn{1}{c}{\textbf{$\mu\pm\sigma$}} & \multicolumn{1}{c}{\textbf{$\mu\pm\sigma$}} & \multicolumn{1}{c}{\textbf{$\mu\pm\sigma$}} & \multicolumn{1}{c}{\textbf{$\mu\pm\sigma$}} & \multicolumn{1}{c}{\textbf{$\mu\pm\sigma$}} & \multicolumn{1}{c}{\textbf{$\mu\pm\sigma$}}\\ 
\multicolumn{1}{p{2.0cm}}{} & \multicolumn{1}{c}{\textbf{dwarfs}} & \multicolumn{1}{c}{\textbf{giants}} & \multicolumn{1}{c}{\textbf{dwarfs}} & \multicolumn{1}{c}{\textbf{giants}} & \multicolumn{1}{c}{\textbf{dwarfs}} & \multicolumn{1}{c}{\textbf{giants}}\\ 
\hline
\hline  \vspace{-.15cm}\\
$\Delta T_{\rm eff}$(K) &       43.6 $\pm$ 137.3     &   $-$54.3 $\pm$ 91.7 &   $-$15.0 $\pm$ 114.7     &   $-$64.3 $\pm$ 117.2 &   $-$60.4 $\pm$ 112.8 &   21.2 $\pm$ 111.3 \\
\multicolumn{1}{c}{\textbf{overall}} & \multicolumn{2}{c}{~$-$5.4 $\pm$ 126.6} & \multicolumn{2}{c}{~$-$57.4 $\pm$ 118.1} & \multicolumn{2}{c}{~~$-$12.0 $\pm$ 125.5} \vspace{.25cm}\\
$\Delta\log g$(dex)     &       0.01 $\pm$ 0.13~\,     &   0.09 $\pm$ 0.22~\, &   $-$0.03 $\pm$ 0.17~\,     &   $-$0.03 $\pm$ 0.33~\, &   $-$0.02 $\pm$ 0.16~\, &   $-$0.14 $\pm$ 0.22~\, \\
\multicolumn{1}{c}{\textbf{overall}} & \multicolumn{2}{c}{0.05 $\pm$ 0.19} & \multicolumn{2}{c}{$-$0.03 $\pm$ 0.31} & \multicolumn{2}{c}{$-$0.09 $\pm$ 0.20} \vspace{.25cm}\\
$\Delta \rm [M/H]$\tnote{c}     &       0.006 $\pm$ 0.082     &   0.039 $\pm$ 0.121 &   $-$0.020 $\pm$ 0.077     &   $-$0.014 $\pm$ 0.102 &   $-$0.010 $\pm$ 0.080 &   $-$0.107 $\pm$ 0.165 \\
\multicolumn{1}{c}{\textbf{overall}} & \multicolumn{2}{c}{0.023 $\pm$ 0.105} & \multicolumn{2}{c}{$-$0.015 $\pm$ 0.099} & \multicolumn{2}{c}{$-$0.072 $\pm$ 0.170} \vspace{.25cm}\\
$\Delta \rm [\alpha/H]$(dex)    &       $-$0.014 $\pm$ 0.082     &   0.005 $\pm$ 0.107 &   ...~~~~~     &   ...~~~~~ &   ...~~~~~ &   ...~~~~~ \\
\multicolumn{1}{c}{\textbf{overall}} & \multicolumn{2}{c}{$-$0.005 $\pm$ 0.096~~~} & \multicolumn{2}{c}{...} & \multicolumn{2}{c}{...} \vspace{.25cm}\\
$\Delta v_{\rm micro}$(km/s)    &       $-$0.34 $\pm$ 0.42~\,     &   $-$0.03 $\pm$ 0.36~\, &   $-$0.44 $\pm$ 0.37~\,     &   $-$0.09 $\pm$ 0.35~\, &   $-$0.10 $\pm$ 0.23~\, &   $-$0.18 $\pm$ 0.19~\, \\
\multicolumn{1}{c}{\textbf{overall}} & \multicolumn{2}{c}{$-$0.18 $\pm$ 0.43~~~} & \multicolumn{2}{c}{$-$0.16 $\pm$ 0.38} & \multicolumn{2}{c}{$-$0.14 $\pm$ 0.22} \vspace{.25cm}\\
$\Delta v_{\rm rad}$(km/s)      & $-$0.01 $\pm$ 4.10~\,     &   $-$0.04 $\pm$ 2.81~\, &   0.49 $\pm$ 1.26~\,     &   0.09 $\pm$ 2.69~\, &   0.69 $\pm$ 4.37~\, &   $-$0.21 $\pm$ 4.09~\,  \\
\multicolumn{1}{c}{\textbf{overall}} & \multicolumn{2}{c}{$-$0.02 $\pm$ 3.51~~~} & \multicolumn{2}{c}{~~0.15 $\pm$ 2.54} & \multicolumn{2}{c}{~~0.11 $\pm$ 4.18} \\ \hline \vspace{-.3cm}
\label{sum_apogalges2}
\end{tabular}
\begin{tablenotes}
     \item[a] The number of common stars refers to the data sets before (after) the cuts. \\
     \item[b] For more information about each filtering criterion, see Sect.~\ref{data}. \\
     \item[c] For APOGEE we use [M/H] rather than [Fe/H].
\end{tablenotes}
\end{threeparttable}
\end{table*}
\end{scriptsize}

\begin{scriptsize}
\begin{table*}
\begin{threeparttable}
\caption{Comparison of the elemental abundance values from the surveys. The average and scatter of abundance differences are calculated after separating MS and GB stars. No data is given in the table cell if there is no available corresponding data. $\Delta$[X/Fe] and $\Delta$A(X) are not included, because it would not essentially affect the trends of differences. The confidence interval is $\pm 1 \sigma$.}
\begin{tabular}{lrrrrrr}
 \hline\hline & 
\multicolumn{2}{c}{\textbf{APOGEE-GALAH}} & 
\multicolumn{2}{c}{\textbf{APOGEE-GES}} &
\multicolumn{2}{c}{\textbf{GALAH-GES}} \\
\hline
\multicolumn{1}{p{2.0cm}}{} & \multicolumn{1}{c}{\textbf{$\mu\pm\sigma$}} & \multicolumn{1}{c}{\textbf{$\mu\pm\sigma$}} & \multicolumn{1}{c}{\textbf{$\mu\pm\sigma$}} & \multicolumn{1}{c}{\textbf{$\mu\pm\sigma$}} & \multicolumn{1}{c}{\textbf{$\mu\pm\sigma$}} & \multicolumn{1}{c}{\textbf{$\mu\pm\sigma$}}\\ 
\multicolumn{1}{p{2.0cm}}{} & \multicolumn{1}{c}{\textbf{dwarfs}} & \multicolumn{1}{c}{\textbf{giants}} & \multicolumn{1}{c}{\textbf{dwarfs}} & \multicolumn{1}{c}{\textbf{giants}} & \multicolumn{1}{c}{\textbf{dwarfs}} & \multicolumn{1}{c}{\textbf{giants}}\\\hline
\hline  
$\Delta \rm [Li/H]$	&	...~~~~~     &   ...~~~~~ &   ...~~~~~     &   ...~~~~~ &   {$-$0.126} $\pm$ 0.286 &   {0.062} $\pm$ 0.255  \\
$\Delta \rm [C/H]$	&	$-$0.031 $\pm$ 0.124     &   ...~~~~~ &   {$-$0.025} $\pm$ 0.145     &   {0.006} $\pm$ 0.200 &   {0.037} $\pm$ 0.136 &   {$-$0.031} $\pm$ 0.133 \\
$\Delta\rm [N/H]$	&	...~~~~~     &   ...~~~~~ &   {0.019} $\pm$ 0.057     &   {$-$0.020} $\pm$ 0.143 &   ...~~~~~ &   ...~~~~~ \\
$\Delta \rm [O/H]$	&	0.027 $\pm$ 0.198     &   $-$0.147 $\pm$ 0.210 &     {$-$0.064} $\pm$ 0.210     &   {$-$0.202} $\pm$ 0.139 &   {$-$0.135} $\pm$ 0.148 &   {$-$0.032} $\pm$ 0.203  \\
$\Delta \rm [Na/H]$	&	$-$0.094 $\pm$ 0.278     &   $-$0.074 $\pm$ 0.225 &   {$-$0.163} $\pm$ 0.281     &   {$-$0.129} $\pm$ 0.212 &   {$-$0.033} $\pm$ 0.088 &   {$-$0.092} $\pm$ 0.176  \\
$\Delta \rm [Mg/H]$	&	$-$0.064 $\pm$ 0.117     &   $-$0.016 $\pm$ 0.143 &   {$-$0.047} $\pm$ 0.094     &   {$-$0.077} $\pm$ 0.097 &   {$-$0.005} $\pm$ 0.113 &   {$-$0.120} $\pm$ 0.228   \\
$\Delta \rm [Al/H]$	&	0.082 $\pm$ 0.123     &   0.052 $\pm$ 0.098 &    {0.073} $\pm$ 0.081     &   {$-$0.010} $\pm$ 0.128 &   {$-$0.033} $\pm$ 0.118 &   {$-$0.032} $\pm$ 0.103  \\
$\Delta \rm [Si/H]$	&	0.006 $\pm$ 0.090     &   $-$0.005 $\pm$ 0.091 &   {0.057} $\pm$ 0.076     &   {$-$0.022} $\pm$ 0.124 &   {0.042} $\pm$ 0.078 &   {$-$0.065} $\pm$ 0.163  \\
$\Delta \rm [S/H]$	&	...~~~~~     &   ...~~~~~ &   {$-$0.007} $\pm$ 0.284     &   {$-$0.047} $\pm$ 0.432   &   ...~~~~~ &   ...~~~~~ \\
$\Delta \rm [K/H]$	& $-$0.114 $\pm$ 0.174     &   0.025 $\pm$ 0.252 &   ...~~~~~     &   ...~~~~~ &   ...~~~~~ &   ...~~~~~ \\
$\Delta \rm [Ca/H]$	&	0.019 $\pm$ 0.126     &   $-$0.009 $\pm$ 0.150 &   {0.086} $\pm$ 0.097     &   {0.016} $\pm$ 0.165 &   {0.002} $\pm$ 0.155 &   {0.022} $\pm$ 0.219  \\
$\Delta \rm [Sc/H]$	&   ...~~~~~	&  ...~~~~~  & ...~~~~~  & ...~~~~~  &  0.105 $\pm$ 0.119 &   $-$0.050 $\pm$ 0.171     \\
$\Delta\rm [Ti/H]$	&	$-$0.155 $\pm$ 0.264     &  $-$0.007 $\pm$ 0.194 &  {$-$0.084} $\pm$ 0.245     &   {$-$0.049} $\pm$ 0.199 &   {0.040} $\pm$ 0.140 &   {$-$0.108} $\pm$ 0.230 \\
$\Delta\rm [Ti~II/H]$	&   $-$0.223 $\pm$ 0.326    &   0.058 $\pm$ 0.235  &   {$-$0.257} $\pm$ 0.319     &   {$-$0.008} $\pm$ 0.194 &   {$-$0.081} $\pm$ 0.119 &   {$-$0.153} $\pm$ 0.265 \\
$\Delta\rm [V/H]$	&	$-$0.053 $\pm$ 0.255     &   $-$0.268 $\pm$ 0.311 &    0.010 $\pm$ 0.249     &   $-$0.133 $\pm$ 0.243 &   0.036 $\pm$ 0.219 &   0.120 $\pm$ 0.325  \\
$\Delta\rm [Cr/H]$	&	0.056 $\pm$ 0.229     &   0.032 $\pm$ 0.183 &  {$-$0.018} $\pm$ 0.219     &   {0.005} $\pm$ 0.196 &   {$-$0.063} $\pm$ 0.117 &   {$-$0.053} $\pm$ 0.171  \\
$\Delta\rm [Mn/H]$	&	$-$0.126 $\pm$ 0.111     &   $-$0.083 $\pm$ 0.198 &   $-$0.153 $\pm$ 0.142     &   $-$0.084 $\pm$ 0.187 &   $-$0.030 $\pm$ 0.112 &   $-$0.034 $\pm$ 0.241  \\
$\Delta\rm [Co/H]$	&	$-$0.217 $\pm$ 0.331     &   0.004 $\pm$ 0.195 &   0.029 $\pm$ 0.365     &   $-$0.083 $\pm$ 0.215 &   0.384 $\pm$ 0.467 &   $-$0.087 $\pm$ 0.198  \\
$\Delta\rm [Ni/H]$	&	0.046 $\pm$ 0.127     &   $-$0.030 $\pm$ 0.133 &   {0.018} $\pm$ 0.084     &   {$-$0.047} $\pm$ 0.138 &   {$-$0.035} $\pm$ 0.144 &   {$-$0.029} $\pm$ 0.159  \\
$\Delta\rm [Cu/H]$	&	...~~~~~      &...~~~~~     & ...~~~~~   &   ...~~~~~  &   0.062 $\pm$ 0.160 &   $-$0.109 $\pm$ 0.248  \\
$\Delta\rm [Zn/H]$	&	...~~~~~     &   ...~~~~~ &   ...~~~~~     &   ...~~~~~ &   0.068 $\pm$ 0.133 &   $-$0.023 $\pm$ 0.272 \\
$\Delta\rm [Zr/H]$	&	...~~~~~  & ...~~~~~  &     ...~~~~~  &    ...~~~~~ &  {0.146} $\pm$ 0.139 &   {0.085} $\pm$ 0.367     \\
$\Delta\rm [Mo/H]$	&	...~~~~~  & ...~~~~~  &    ...~~~~~   & ...~~~~~    &    ...~~~~~ &   0.058 $\pm$ 0.375    \\
$\Delta\rm [Ba~II/H]$	&	...~~~~~     &   ...~~~~~ &   ...~~~~~     &   ...~~~~~ &   {$-$0.009} $\pm$ 0.205 &   {0.138} $\pm$ 0.359  \\
$\Delta\rm [La~II/H]$	&	...~~~~~     &   ...~~~~~ &   ...~~~~~     &   ...~~~~~ &   {0.271} $\pm$ 0.214 &   {$-$0.140} $\pm$ 0.216  \\
$\Delta\rm [Ce~II/H]$	&	...~~~~~     &   ...~~~~~ &   ...~~~~~     &   ...~~~~~ &   {0.104} $\pm$ 0.220 &   {$-$0.196} $\pm$ 0.224 \\
$\Delta\rm [Nd~II/H]$	&	...~~~~~     &   ...~~~~~ &   ...~~~~~     &   ...~~~~~ &   {0.315} $\pm$ 0.219 &   {0.052} $\pm$ 0.212 \\
$\Delta\rm [Eu~II/H]$	&	...~~~~~     &   ...~~~~~ &   ...~~~~~     & ...~~~~~     & ...~~~~~     &   {$-$0.133} $\pm$ 0.178   \\ \hline \vspace{-.3cm}
\label{sum_apogalges3}
\end{tabular}
\end{threeparttable}
\end{table*}
\end{scriptsize}

\begin{scriptsize}
\begin{table}
\begin{threeparttable}
\caption{Overall average of the reported individual parameter errors from the golden samples.}
\begin{tabular}{lcccc}
\hline\hline & 
\multicolumn{1}{c}{$\sigma_{v_{\rm rad}}$} & 
\multicolumn{1}{c}{{$\sigma_{T_{\rm eff}}$}} &
\multicolumn{1}{c}{{$\sigma_{\log~g}$}} &
\multicolumn{1}{c}{$\sigma_{\rm [Fe/H]}$\tnote{a}} \\ \hline
\textbf{APOGEE} & $0.04$ km/s & $14.9$ K & $0.03 
$ dex & $0.008$ dex\\
\textbf{GALAH} & $0.11$ km/s & $92.3$ K & $0.21
$ dex & $0.077$ dex\\
\textbf{GES} & $0.29$ km/s & $60.6$ K & $0.11
$ dex & $0.060$ dex \\ \hline \vspace{-.3cm}
\label{err}
\end{tabular}
\begin{tablenotes}
     \item[a] For APOGEE we use [M/H] rather than [Fe/H].
\end{tablenotes}
\end{threeparttable}
\end{table}
\end{scriptsize}

\begin{table}
\begin{threeparttable}
\caption{Median absolute deviation  of the reported parameters and abundances from the golden samples.}
\begin{tabular}{lccc}
\hline \hline  & 
\multicolumn{3}{c}{\textbf{APOGEE$-$GALAH}} \\ \hline
 & \textbf{~~dwarfs~~} & \textbf{~~giants~~} & \textbf{~~overall~~} \\ \hline \hline  \vspace{-.2cm}\\
$\Delta T_{\rm eff}$ (K) \hspace{1.4cm} & 78.4 & 58.3 & 67.1\\
$\Delta \log~g$ (dex) & $0.07$ & $0.14$ & 0.10\\
$\Delta$[Fe/H]\tablefootmark{a} (dex) & 0.043 & $0.065$ & 0.052 \\
$\Delta[\alpha\rm /Fe]$ (dex) & 0.038 & 0.053 & 0.045 \\
$\Delta v_{\rm micro}$ (km/s) & 0.47 & 0.20 & 0.30 \\
$\Delta v_{\rm rad}$ (km/s) & 0.20 & 0.21 & 0.20  \\
\hline \hline 
& \multicolumn{3}{c}{\textbf{APOGEE$-$GES}} \\
\hline 
& \textbf{~~dwarfs~~} & \textbf{~~giants~~} & \textbf{~~overall~~} \\ 
\hline \hline  
$\Delta T_{\rm eff}$ (K) & 52.8 & 71.4 & 66.6\\
$\Delta \log~g$ (dex) & 0.08 & 0.14 & 0.13 \\
$\Delta$[Fe/H]\tnote{a} (dex) & 0.035 & 0.058 & 0.053 \\
$\Delta v_{\rm micro}$ (km/s) & 0.51 & 0.20 & 0.22 \\
$\Delta v_{\rm rad}$ (km/s) & 0.51 & 0.25 & 0.27 \\
\hline \hline 
& \multicolumn{3}{c}{\textbf{GALAH$-$GES}} \\
\hline 
& \textbf{~~dwarfs~~} & \textbf{~~giants~~} & \textbf{~~overall~~} \\ 
\hline \hline 
$\Delta T_{\rm eff}$ (K) & 88.9 & 65.7 & 76.2 \\
$\Delta \log~g$ (dex) & 0.09 & 0.14 & 0.12 \\
$\Delta$[Fe/H] (dex) & 0.050 & 0.097 & 0.069 \\
$\Delta v_{\rm micro}$ (km/s) & 0.14 & 0.20 & 0.16 \\
$\Delta v_{\rm rad}$ (km/s) & 0.57 & 0.27 & 0.36 \\ \hline \vspace{-.3cm}
\label{mad_param}
\end{tabular}
\begin{tablenotes}
     \item[a] For APOGEE we use [M/H] rather than [Fe/H].
\end{tablenotes}
\end{threeparttable}
\end{table}

The average difference in RV is $0.15\pm2.54$~km/s for the APOGEE$-$GES catalog, and  $94.3\%$ of stars are within 1$\sigma$ uncertainty (see \autoref{param_drv} middle row). The average and scatter of RV discrepancies are $0.49\pm1.26$~km/s and $0.09\pm2.69$~km/s for dwarfs and giants, respectively. We note that the offset is more significant for MS stars, and is larger than the combined uncertainty of the APOGEE$-$GES catalog, which is $0.28$~km/s. This offset is reduced to $0.42$~km/s when we filter out the dwarfs with a RV difference larger than $\pm5$~km/s. Such large discrepancies may happen when the initial atmospheric parameters used during the cross-correlation have a large offset from the actual parameters of the stars.

In the GALAH$-$GES common sample (\autoref{param_drv}, top row) we found an offset and scatter of $0.11$~km/s and $4.18$~km/s in RV differences, and $92.9\%$ of the considered stars are included in this interval. For dwarfs and giants the discrepancy is $0.69\pm4.37$~km/s and $-0.21\pm4.09$~km/s, respectively. Similarly to the case of APOGEE$-$GES catalog, the offset is more significant for MS stars, which   exceeds the GALAH$-$GES combined estimated precision of $0.30$~km/s. 
This relatively high offset in dwarfs is partially the result of 12 MS stars with $\Delta v_{\rm rad}>5$~km/s in common between GALAH and GES data set. After filtering these stars out the mean is reduced to $0.39$~km/s.

Using Gaia as a reference, \citet{tsantaki_2022} presents a comprehensive catalog of RVs, built by homogeneously merging the RV determinations of the largest ground-based spectroscopic surveys to date, namely  
APOGEE, GALAH, Gaia-ESO, RAVE, and LAMOST. In their work, the uncertainties are normalized using repeated measurements, or the three-cornered hat method, and a cross-calibration of the RVs as a function of the main parameters is included. They estimate the accuracy of the RV zero-point to be about $-0.01$~km/s, $0.04$~km/s, and $0.10$~km/s and the RV precision to be in the range $3.89$~km/s, $3.97$~km/s, and $4.72$~km/s compared to APOGEE DR16, GALAH DR2, and  GES DR3, respectively. While these numbers are in the range of what we find here, an ``apples to apples'' comparison with our values is not possible as we use more recent data releases for all surveys and other factors. 

Overall we find that APOGEE, GALAH, and GES generally measure the same RVs within uncertainties, but a higher than expected scatter around the mean differences is observed. Currently, we do not have an explanation of such large radial velocity differences. 

In the bottom panel of the APOGEE$-$GALAH row in \autoref{param_drv}, one can see a set of stars at particularly high  RVs: $\sim$235~km/s. We conclude that most of these stars are part of the massive globular cluster $\omega$~Cen (NGC 5139), and RV differences show a clear $0.27$~km/s offset in the APOGEE$-$GALAH overlapping data set. For these 130 cluster members, the average RVs are $234.95\pm0.55$~km/s and $234.68\pm0.56$~km/s from APOGEE DR17 and GALAH DR3, respectively. The latter result is closer to the value reported by \citet{baumgardt_2019} as they used \textit{Gaia} DR2 to study the motion and proper kinematics of globular clusters and published a mean RV of 234.28$\pm$0.24~km/s for $\omega$~Cen.

\section{Discussion of differences: Main stellar parameters}\label{disc_diff}

In this section we discuss in detail the discrepancies of the main stellar parameters ($T_{\rm eff}$, $\log g$, metallicity, and $v_{\rm micro}$) derived by the three spectroscopic surveys. We note that the relevant atmospheric parameter (AP) difference, $\Delta \rm (AP)$=\{$\Delta T_{\rm eff}$,$\Delta\log g$,$\Delta$[M/H] or $\Delta$[Fe/H],$\Delta v_{\rm micro}$\}, is calculated as $\rm AP_{\rm(APOGEE)}$$-$$\rm AP_{\rm (GALAH)}$, $\rm AP_{\rm (APOGEE)}$$-$$\rm AP_{\rm(GES)}$, and $\rm AP_{\rm (GALAH)}$$-$$\rm AP_{\rm (GES)}$ for the APOGEE$-$GALAH, APOGEE$-$GES, and GALAH$-$GES catalogs, respectively. The three surveys followed different approaches to determine these parameters, which we briefly outline here before discussing the differences.

\begin{figure}
\centering
\includegraphics[width=.475\textwidth,angle=0]{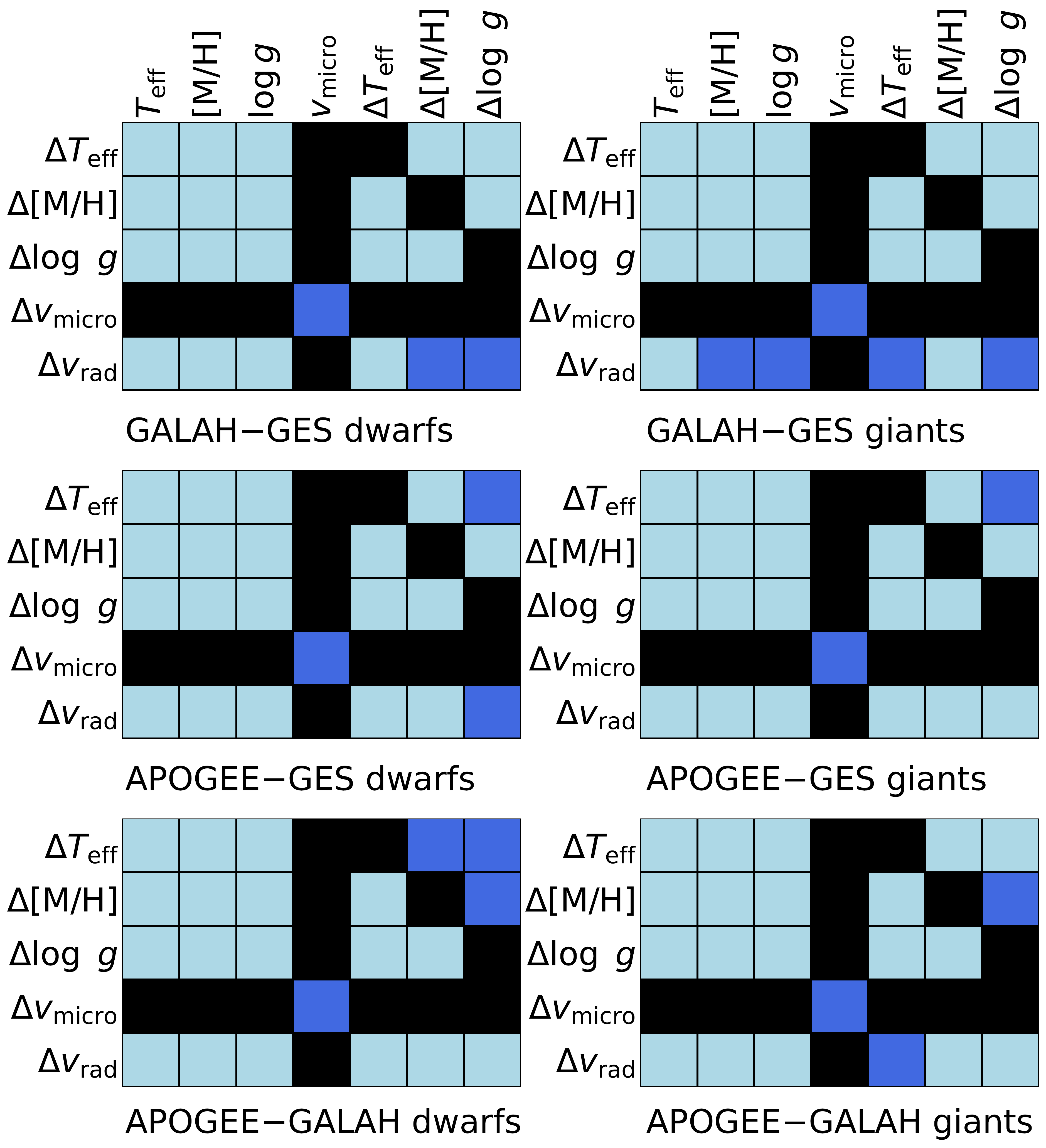}
\caption{
Correlations between the main parameter pairs in the three overlapping catalogs. The strength of the correlations between each pair is color-coded: light blue cells represent no significant correlations ($<0.8\sigma$);    dark blue cells represent weak correlations ($0.8\sigma-1.5\sigma$). Black cells are not relevant due to the absence of reported uncertainties of $v_{\rm micro}$ or because they link the  same quantities (e.g., $\Delta T_{\rm eff}$ with $\Delta T_{\rm eff}$). The estimated strength of the correlations is based on the comparison of the correlation slopes with the average of the reported individual uncertainties of stellar parameters and RVs. The stars have been separated into dwarfs (left column) and giants (right column).}
\label{matrix_param} 
\end{figure}

As described in Sect.~\ref{analysis} above, stellar parameters and abundances for APOGEE DR17 were determined using ASPCAP, which finds the best fit between combined observed visit spectra and synthetic model spectra in APOGEE DR17 \citep{abdurro_2021}. As in previous data releases, the reported individual uncertainties are estimated by analyzing the scatter for stars having repeated observations \citep{abdurro_2021}. Here and in Sect.~\ref{disc_diff_abund} we make use of uncalibrated $T_{\rm eff}$, $\log~g$, $\log(v_{\rm micro})$, [$\alpha$/M], and [M/H] values reported under the ASPCAP output array \texttt{FPARAM} with their associated errors. Similarly to previous data releases, systematic differences between spectra taken from APO and LCO  were corrected by applying the median difference for the stars observed at the two observatories \citep{abdurro_2021}. 

In GALAH DR3 all stellar parameters and elemental abundances were estimated via the spectrum synthesis code Spectroscopy Made Easy \citep[SME, ][]{sme_1996,sme_2017} with 1D MARCS stellar atmosphere models \citep{marcs_2008}.
Tests were performed to validate the obtained parameters in terms of their accuracy and precision \citep{buder_2021}. Accuracy assessment involves commonly used comparison samples: the Sun, \textit{Gaia} FGK benchmark stars \citep[GBS, ][]{heiter_2015b,jofre_2018}, photometric $T_{\rm eff}$ from the  infrared flux method  \citep[IRFM, ][]{casagrande_2010}, stars with asteroseismic data, and cluster members, while internal uncertainty estimations and repeated observations are used for the precision analysis \citep{buder_2021}.  

Based on a common suitable line list, set of model atmospheres, grid of synthetic spectra, and approach to data formats and standards for the analyses, the parameters returned by the various working groups (WGs) contributing to the GES program at the end of the parameter determination are processed 
by the homogenization WG to produce a set of recommended parameters for the survey. The same kind of quality assessment and calibration as used by the GALAH team is applied by the GES team \citep{randich_2022}.
Parameters from the quality GES DR5 catalog are displayed in the third panel of \autoref{teff_logg}, top row.

In order to properly discuss the differences, we calculate the average $T_{\rm eff}$, $\log g$, metallicity uncertainties of those APOGEE DR17, GALAH DR3, and GES DR5 stars that are contained by the relevant filtered catalogs (golden  samples in \autoref{teff_logg}, first row) and use these average uncertainties in the evaluation of the observed discrepancies. 
The mean values of these errors, $\sigma_{T_{\rm eff}}$, $\sigma_{\log~g}$, $\sigma_{\rm [Fe/H]}$ are presented in \autoref{err}.
If the discrepancy distribution is approximately a symmetric Gaussian and there are no systematic effects, we can use the sum of squares of corresponding uncertainty values to compare the observed differences. This is the case if the uncertainties originating from each survey are combined. This also applies to the [X/H] errors calculated from the reported [X/Fe] and [Fe/H] uncertainties. However, it is worth noting that this is an oversimplification in estimating the combined errors as systematic effects always exist between parameters, though their extent is often not known precisely enough to correctly take them into account. 
We also calculate and evaluate the dispersion of the various distributions measured through their median absolute deviation (MAD). For the main parameters, the MAD values are presented in \autoref{mad_param}, and for the abundances  in \autoref{mad_abund}. We conclude that the MAD values are typically $~\sim 40$ to $80$\%  lower than the standard deviation of the differences. 

\autoref{matrix_param} displays the  correlations found between parameter pairs in all three overlapping catalogs. Each cell corresponds to the analysis of a specific parameter plane, and for each parameter pair a linear least-squares fit was made for the set of overlapping stars between the surveys. The degree to which a correlation exists or not is represented by the color-coding. Based on the slope of the fits, denoted by $a$, we deduce the estimated uncertainty of the parameter discrepancies. Cell by cell, this analysis is based on an approximation that the parameter difference is completely caused by the uncertainty of its pair parameter only. Therefore, we define three categories: ($i$) no significant correlation if $a<0.8\sigma$, ($ii$) potential weak correlation if $0.8\sigma<a<1.5\sigma$, ($iii$) substantial correlation if $1.5\sigma<a$.
No substantial correlations (case $iii$) are found between any of the surveys. For the majority of cells we found no statistically relevant correlations (case $i$), implying that the uncertainty is dominated simply by the individual errors determined for the individual derived parameters from each survey.
However, as \autoref{matrix_param} indicates, several weak correlations (case $ii$) exist among the parameters. For example, specifically for the APOGEE$-$GALAH dwarf subsample, there is a weak correlation between $\Delta T_{\rm eff}$ (first row) and $\Delta$[M/H] (sixth column)  implying a $T_{\rm eff}$ discrepancy $0.8-1.5$ times higher than from metallicity uncertainties alone.
In the following subsections we go into the details of the differences in parameters.

\begin{figure*}
\centering
\includegraphics[width=\textwidth,angle=0]{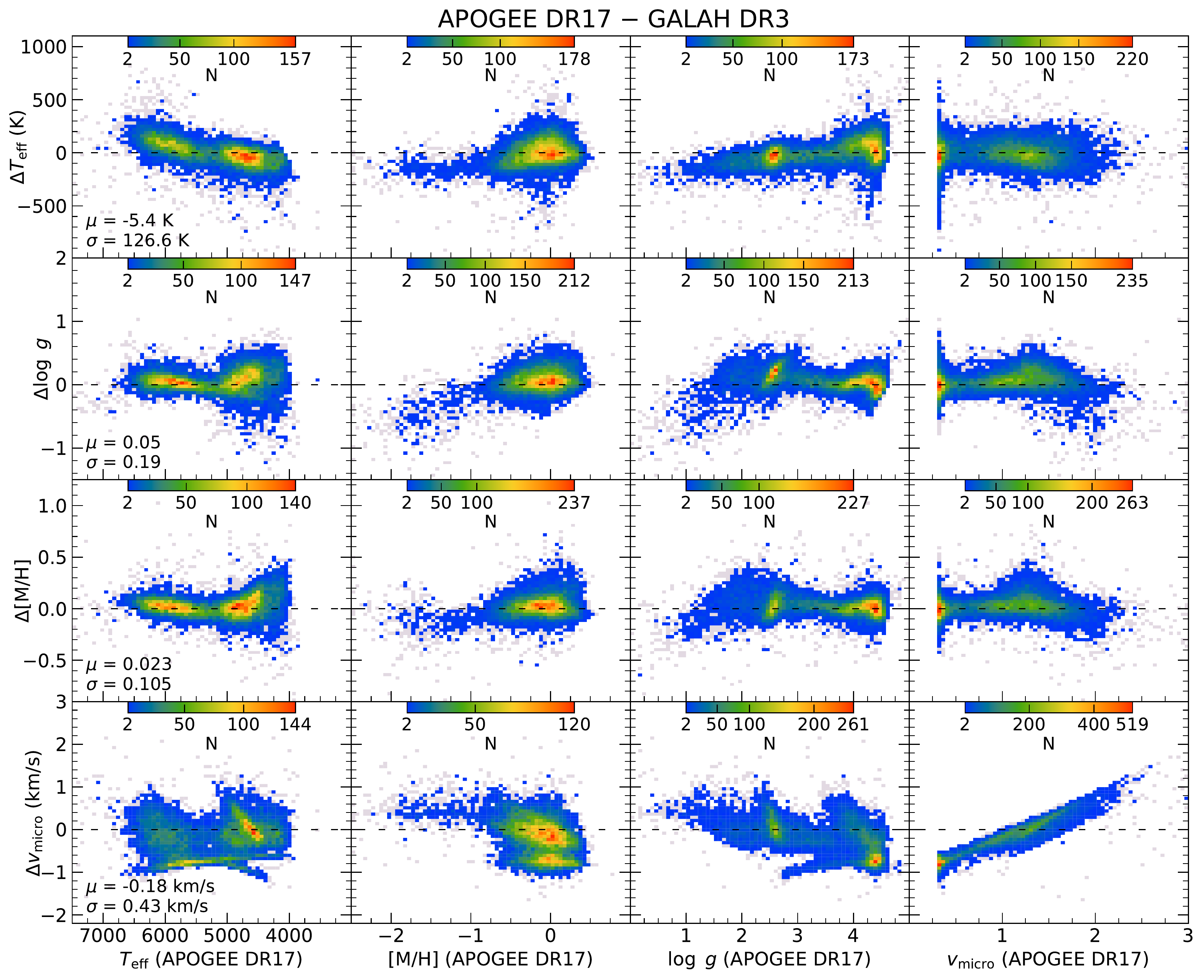}
\caption{Distributions of effective temperature (top row), surface gravity (second row), metallicity (third row), and microturbulence velocity differences (bottom row) as functions of the  main stellar parameters from the APOGEE$-$GALAH overlapping stellar catalog, for dwarfs and giants combined. The number density of stars (starting with two stars) over the parameter space spanned is provided by the color-coding. The binned star densities over the parameter spaces, and the (overall) mean and standard deviation of parameter differences are shown in each panel as well. Bins representing one star are shown in gray. The horizontal axes indicate $T_{\rm eff}$ (K), metallicity, $\log g$ (dex), and $v_{\rm micro}$ (km/s) from APOGEE DR17, and the parameter differences are calculated as $\Delta\rm (AP)=AP_{(APOGEE)}-AP_{(GALAH)}$.}
\label{param_dparam}
\end{figure*}

\subsection{Effective temperature}\label{disc_teff}

Effective temperature is arguably the most important main stellar parameter that has the largest effect on the structure of  stellar atmosphere   and the  formation of absorption lines. Similar to the RV analysis performed earlier, we compare the mean differences and scatters to the average of the reported individual errors combined for the relevant survey pairs.

Discrepancies in $T_{\rm eff}$ are shown in the top row of \autoref{param_dparam} for the APOGEE$-$GALAH overlapping globally filtered sample. The overall statistics, including the mean offset and standard deviation of $T_{\rm eff}$ separated for MS and RGB stars, are listed in \autoref{sum_apogalges2}. That table also contains the calculated differences for the APOGEE$-$GES and GALAH$-$GES common stars.

In the APOGEE$-$GALAH quality data set, we find a discrepancy of $-5.4\pm126.6$~K for $T_{\rm eff}$, as shown in \autoref{param_dparam}. While 78.2\% of these stars are located within the  $\pm1\sigma$ range, there is also a linear trend in the $T_{\rm eff}-\Delta T_{\rm eff}$ plane. 
The offset is significantly different for dwarfs ($43.6 \pm 137.3$~K) and giants ($-54.3 \pm 91.7$~K), as seen in \autoref{sum_apogalges2}, but the collective offset spans the interval of predicted uncertainties of $\sigma_{T_{\rm eff}}=91.5$~K propagated from the APOGEE and GALAH golden samples. Separately, the offsets still remain lower than the preliminary calculated APOGEE$-$GALAH sample uncertainty of 93.3~K and 94.6~K for MS and RGB stars, respectively. The GALAH team reported a trend toward increasingly underestimated $T_{\rm eff}$ for hotter dwarfs, similar to previous GALAH analyses \citep{buder_2021}. This is in a good agreement with the positive offset of MS presented here. 

We note that the issue of altering the sign of the offsets is visible in the $\Delta  T_{\rm eff}$ plot as a function of $T_{\rm eff}$ and $\log g$ in the relevant panels of \autoref{param_dparam}. We do not  find 
a symmetric distribution or a linear trend in the [M/H] versus $\Delta T_{\rm eff}$ plot, but a systematic deviation for metal-poor stars is apparent. Lastly, we find no notable correlation between microturbulence and $\Delta T_{\rm eff}$.
In addition to the overall analysis of the stellar distributions in \autoref{param_dparam}, the bottom row of \autoref{matrix_param} depicts the following considerations: for dwarfs the correlations of $\Delta T_{\rm eff}$  with $\Delta$[M/H] and $\Delta\log g$ remain weak, within the determined uncertainties, and in the other cases and for the giants, $T_{\rm eff}$ remains within the combined uncertainty ranges determined for the APOGEE$-$GALAH catalog.

In APOGEE$-$GES catalog we observe an offset in $T_{\rm eff}$ of about $-57$~K together with a scatter of $118$~K; 70.8\% of these stars cover the variation range of $\pm 1 \sigma$. When looking at the difference between the dwarfs and giants, we find that the mean $T_{\rm eff}$ discrepancy for both MS and RGB stars are negative, namely $-15.0 \pm 114.7$~K and $-64.3 \pm 117.2$~K, respectively (see \autoref{sum_apogalges2}); the GES temperatures are generally hotter than those determined by APOGEE by the average amounts noted. These offsets are not significantly higher than the expected combined $\sigma_{T_{\rm eff}} = 53.1$~K precision that is calculated for the APOGEE and GES quality stars.

We also present the $T_{\rm eff}$ differences between GALAH and GES with respect to the main stellar parameters. The overall offset and scatter is $-12.0\pm125.5$~K for our filtered data set of common targets, and 75.3\%  of the sample falls in this range, while the separate results presented in \autoref{sum_apogalges2} reveal that the effects of different measurements and/or derivations play a role in dwarfs and giants. Although neither of the offsets is higher than the average uncertainty of the propagated errors ($\sigma_{T_{\rm eff}}=100.3$~K), it is interesting that the discrepancy seems more notable for dwarfs ($-$60.4~K) than for giants (21.2~K). However, there are relatively few objects in the MS parameter space (at high $\log g$ and $T_{\rm eff}$) to perform  meaningful statistics (see \autoref{teff_logg}). 

We conclude that the three high-resolution spectroscopic programs measure similar $T_{\rm teff}$ values for the quality samples within the determined uncertainty of each survey. The previously found systematic difference between APOGEE spectroscopic and photometric $T_{\rm eff}$ can be seen in the APOGEE$-$GALAH comparison as well.

\subsection{Surface gravity}\label{disc_logg}

In this section we discuss the differences in surface gravity as a function of the stellar parameters,   as above for $T_{\rm eff}$ (see Sect.~\ref{common_descr}). 

The second row in \autoref{param_dparam} shows the $\log g$ discrepancies with respect to the measured stellar parameters for the APOGEE$-$GALAH common stars. The mean offset and scatter calculated for MS and RGB stars are listed in \autoref{sum_apogalges2}. Comparing the APOGEE $\log g$ with GALAH is interesting because APOGEE measures it from fitting the spectra, while GALAH updated the spectroscopic surface gravities with isochrones \citep{buder_2021}. 

In the APOGEE$-$GALAH catalog we see an overall offset of $0.05$~dex with a scatter of $0.19$~dex, and 77.5\% of these stars belong to this uncertainty range.
In terms of dwarfs, the discrepancy is $0.01\pm0.13$~dex, and we find an interesting curving downturn in the $\log g-\Delta \log g$ parameter space around dwarfs, although a slight offset and a weak linear rising trend at the cool end exist with respect to $T_{\rm eff}$. 
We can see a larger mean difference and standard deviation for giants, including the red clump, which is $0.09\pm0.22$~dex. APOGEE seems to have derived systematically higher $\log g$ for giants than GALAH, and this is also stated in \citet{holtzman_2018}:  APOGEE ASPCAP (raw) spectroscopic $\log g$ values for giants are systematically higher than those derived from asteroseismology. Instead,  \citet{desilva_2015} found good agreement between the GALAH's and asteroseismic $\log g$ values.  Therefore, calibration was necessary in APOGEE (see Sect.~\ref{calib}). Comparing calibrated results from APOGEE DR14 and uncalibrated values from LAMOST DR3, \citet{anguiano_2018} also found a positive deviation of $0.14$~dex with a scatter of $0.25$~dex in $\log g$. We note that the metal-poor giant members of $\omega$~Cen account for the significant $\Delta \log g < -0.4$~dex values shown in the second panel from the left in the second row of \autoref{param_dparam}.

The calculated uncertainty of $\sigma_{\log g}=0.21$~dex propagated from APOGEE and GALAH is always smaller than the discrepancy seen for the MS and the RGB stars. Moreover, \autoref{matrix_param} (third row) indicates no significant correlation between the main parameter errors and the differences between the APOGEE$-$GALAH parameters. 
The $\log~g$ discrepancy between APOGEE and GALAH is large at very low metallicities (third panel from the left in the second row); however, the spectroscopic surface gravities for cooler dwarfs tend to be underestimated \citep{holtzman_2018}, which is also found within our analysis. \citet{nandakumar_2020} also found systematically higher $\log g$ values for APOGEE DR16 than for GALAH SME, in agreement with our results. The pattern of discrepancy drawn by red clump stars ($\log g \sim 2.5$~dex) and MS stars is retained as well. 

In APOGEE$-$GES a mean discrepancy of $-0.03$~dex and a scatter of $0.31$~dex is calculated. The $\pm 1\sigma$ interval covers 75.6\% of the stars in common. Separating the MS and RGB stars, we find a $-0.03\pm0.17$~dex average $\log g$ difference for dwarfs and $-0.03\pm0.33$~dex for giants (see \autoref{sum_apogalges2}). 
The APOGEE$-$GES combined uncertainty is estimated as $\sigma_{\log g}=0.12$~dex, which is larger than the offset observed. No clear trends can be observed in the discrepancy distributions.

The $\log g$ discrepancies as functions of stellar parameters of stars included in the GALAH$-$GES quality catalog follow. We find that the mean deviation of the measured $\log g$ values is $-0.09$~dex with a scatter of $0.20$~dex, and 75.3\% of the sample is located in the $\pm 1\sigma$ range. As listed in \autoref{sum_apogalges2}, the offset is $-0.02\pm0.16$~dex for dwarfs and a more dominant offset of $-0.14\pm0.22$~dex for giants can be seen. Similarly to the APOGEE$-$GES data set, the majority of the stars is located on the RGB. 
While GALAH$-$GES giants have the largest average $\log g$ difference among all the discussed data sets, this discrepancy stays within the combined internal precision of $\sigma_{\log g} = 0.24$~dex. Finally, no systematic trends in $\Delta\log g$ can be found with respect to the main parameters.

As a result, no significant surface gravity offset and no substantial correlation beyond the average (and combination) of individual parameter uncertainties are found. APOGEE, GALAH, and GES determine their $\log g $ values within their estimated uncertainties.

\subsection{Metallicity}\label{disc_met}

Here  we discuss the differences between the derived metallicity values, which is an important tracer of stellar populations. Similarly to the previous main parameters, the discrepancy is compared to the combined estimated uncertainties.

\begin{figure}
\includegraphics[width=.475\textwidth,angle=0]{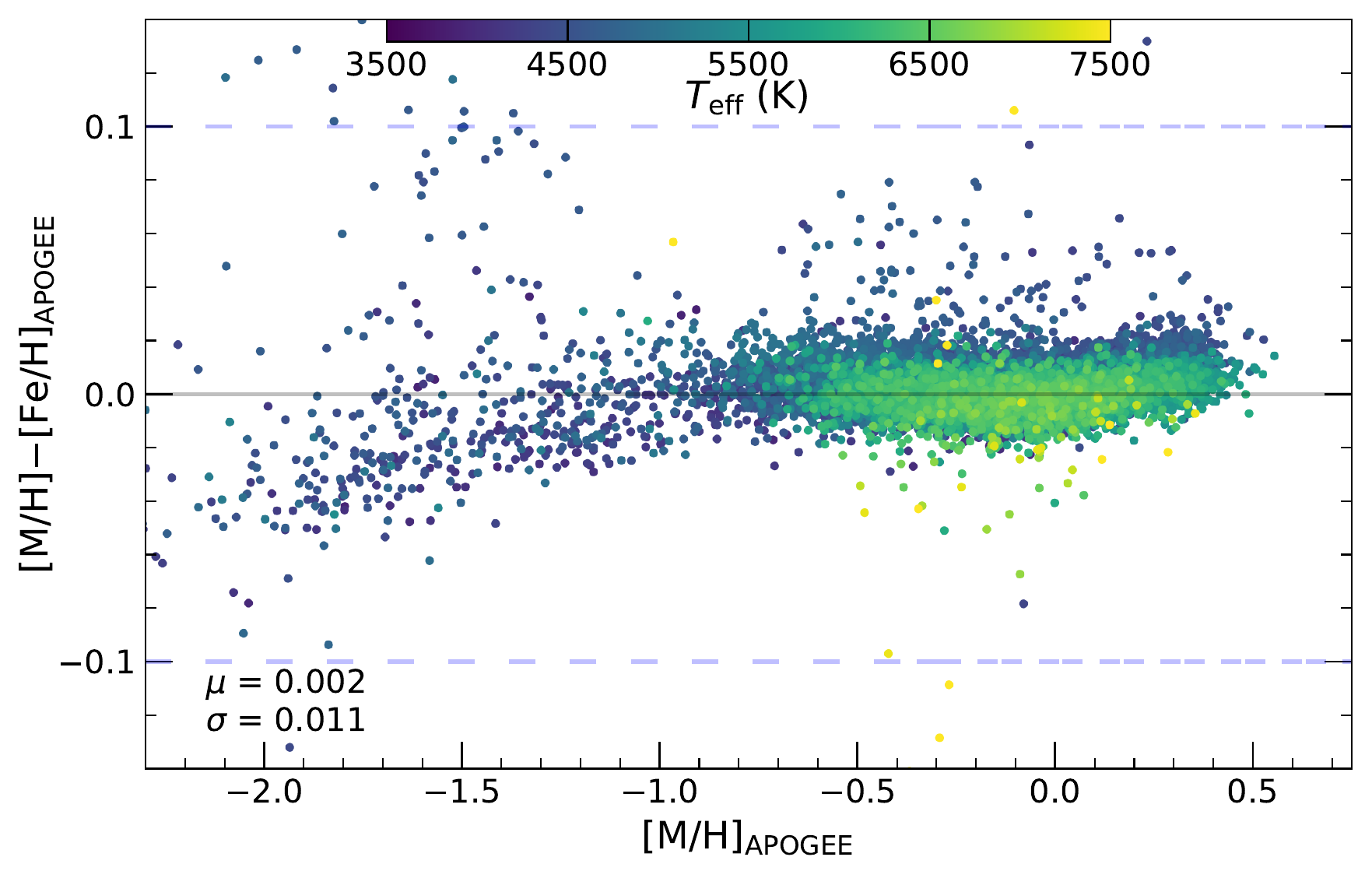}
\caption{Overall metallicity vs. the difference between overall metallicity and iron abundance  color-coded by $T_{\rm eff}$ published in APOGEE DR17. The indicated stars are from the APOGEE$-$GALAH common stellar catalog and global cuts are applied as well. The dashed lines indicate a difference of $\pm0.1$~dex (expected uncertainty from abundance analysis).}
\label{met_feh}
\end{figure}

In APOGEE, the overall metallicity is derived by ASPCAP fitting all spectral lines globally, not only by lines of iron  \citep{garcia_2016}. However, it is still a valid decision to consider that  [M/H] and [Fe/H] are  equivalent, as the number of iron lines in the spectra of FGK-type stars is by far the largest among all elements, so overall metallicity is essentially the iron abundance within the margin of precision of $\pm0.1$~dex \citep{jofre_2019}. 
\autoref{met_feh} depicts the differences between uncalibrated [M/H] and [Fe/H] for stars included in the APOGEE$-$GALAH quality catalog. While we can find a higher scatter at low metallicities (and at the highest $T_{\rm eff}$ values), the average discrepancy and scatter is 0.002$\pm$0.011~dex, which lies within $0.1$~dex (denoted by dashed lines in \autoref{met_feh}). When discussing the precision of abundances we use this $0.1$~dex in Sects.~\ref{disc_met} and \ref{disc_diff_abund} as a guide of an indicative precision that we expect the high-resolution surveys can reach for elements with well-defined absorption lines.

Furthermore, \citet{garcia_2016} reported that uncertainties tend to be more significant  toward low metallicities and temperatures, which we also experience here. Therefore, we consider overall metallicity and iron abundance equivalent in APOGEE DR17, and we use [M/H] in the comparisons.

For all stars included in the APOGEE$-$GALAH overlapping quality catalog, the published metallicity discrepancies with respect to main parameters are represented in the third row of \autoref{param_dparam}. We find that the mean metallicity difference is $0.023$~dex with a scatter of $0.105$~dex between the stars collectively, and 78.9\% of the samples ranges in the $\pm1\sigma$ interval. In the [M/H]$-\Delta$[M/H] and $v_{\rm micro}-\Delta$[M/H] parameter spaces, it is clearly recognized that APOGEE systematically measures higher metallicity values.

Specifically for the MS stars, we find an offset of $0.006$~dex along with a $0.082$~dex scatter. As shown in the relevant (third row) $T_{\rm eff}-\Delta$[M/H] and $\log g-\Delta$[M/H] planes in \autoref{param_dparam}, there is a larger mean and standard deviation ($0.039\pm0.121$~dex) in $\Delta$[M/H] for RGB stars (see \autoref{sum_apogalges2}), and the offset is strongly dependent on the $\log~g$ differences between APOGEE and GALAH (\autoref{matrix_param}). This may suggest that systematic offsets in surface gravity may translate to discrepancies in metallicity. 
The APOGEE$-$GALAH combined internal uncertainty is $\sigma_{\rm [M/H]}=0.077$~dex, we conclude that the derived mean discrepancies stay within this range. Therefore, an overall agreement is found in the measured metallicities between APOGEE DR17 and GALAH DR3, but users should be aware that the $\Delta$[M/H] is larger at low $T_{\rm eff}$ values.

\citet{soubiran_2022} performed the assessment of [Fe/H] determinations of FGK stars in
spectroscopic surveys. In addition to  conducting several tests, they looked for trends in the distribution of residuals versus Gaia G magnitude and $T_{\rm eff}$, $\log~g$, and [M/H]. In general, they found a consistency between APOGEE DR16 (calibrated values),  GALAH DR3, and  GES DR3. The offsets and scatters are on the same order as in our results, and their plotted distribution patterns also correspond to those in \autoref{param_dparam}.

\citet{anguiano_2018} also observed a small offset (between APOGEE and LAMOST) in calibrated [Fe/H] of about $0.06$~dex together with a scatter of $0.13$~dex, and also discovered a more significant discrepancy for cooler stars.
Furthermore, when comparing GALAH DR3 with GALAH DR2, \citet{buder_2021} also found a trend of underestimated [Fe/H] for the metal-rich giants and red clump stars, which suggests that their synthetic spectra have a slight inaccuracy \citep{buder_2021}. The difference for all stars with unflagged abundances between APOGEE DR16 and GALAH shows a slightly lower [Fe/H] for GALAH ($-0.05\pm0.14$~dex) \citep{buder_2021}. According to our study, this underestimation is translated into the positive offset between APOGEE and GALAH.

The main parameter dependences of $\Delta$[Fe/H] for the APOGEE$-$GES quality stellar data set is discussed in the next step. The average difference is $-0.015$~dex with a scatter of $0.099$~dex, and 76.4\% of the stars is located within this uncertainty range. As the vast majority of these stars are cooler than $5500$~K and have $\log g<$3.75~dex, stars contributing to the statistics are mostly giants. The discrepancy is $-0.020\pm0.077$~dex and $-0.014\pm0.102$~dex for MS and RGB stars, respectively.
The average reported error is $\sigma_{\rm [M/H]}=0.060$~dex in the combined APOGEE and GES quality samples. In conclusion, the offset is small without any remarkable systematic trends.

In terms of the GALAH$-$GES overlapping stars, an overall combined uncertainty of $\sigma_{\rm [Fe/H]}=0.097$~dex is calculated, which suggests that the present offset of $-0.107\pm0.165$~dex for giants is significant. Since the number of the comprised GALAH$-$GES stars is relatively small, we may conclude that the large offset may be the result of the small sample size.

In summary, for  the metallicity discrepancies we find no significant offsets and correlations of main parameters beyond the expected uncertainties, except for the mean discrepancy being larger than the expected uncertainty within GALAH$-$GES and a weak linear relation between $\Delta\log g$ and $\Delta$[M/H] in the case of APOGEE$-$GALAH. We note that stars with $T_{\rm eff} < $ 4500~K have larger metallicity differences than the rest of the stars in the APOGEE$-$GALAH common sample.

\subsection{Microturbulent velocity}\label{disc_vmicro}

Here we present a discussion of the empirical (line broadening) atmosphere parameter $v_{\rm micro}$. Microturbulent velocity strongly influences the derivation of abundances, so it is important to assess its accuracy among the sky surveys. 

The $v_{\rm micro}$ discrepancy distribution for the overlapping quality catalog APOGEE$-$GALAH is presented in the panels of \autoref{param_dparam} in the bottom row. The discrepancy ranges from $\sim-1.1$~km/s to $+1.5$~km/s and we observe an average difference of $-0.18$~km/s with a standard deviation of $0.43$~km/s, and specifically $-0.34\pm0.42$~km/s and $-0.03\pm0.36$~km/s for MS and RGB stars (see \autoref{sum_apogalges2}). In the $\log g$ versus $\Delta v_{\rm micro}$ plane it is clear that we have a significant negative offset at high $\log g$ values, suggesting that  the two surveys use vastly different $v_{\rm micro}$ for the dwarf stars. In addition to  the strong correlation with respect to $v_{\rm micro}$ (last panel), further interesting patterns appear in the $T_{\rm eff}-\Delta v_{\rm micro}$ and [M/H]$-\Delta v_{\rm micro}$ planes, for example a strong negative $\Delta v_{\rm micro}$ gradient as a function of $T_{\rm eff}<5000$~K, and we find detached groups of stellar densities near  solar-like metallicities, which are the MS stars. 
The substantial correlation with respect to $v_{\rm micro}$ is also represented in \autoref{matrix_param}. Here we fit $\Delta v_{\rm micro}$ as a linear function of $v_{\rm micro}$, and find a slope of $\sim0.9$ which means that the entire difference is comparable to the combined uncertainties in the APOGEE$-$GALAH sample. 

The two surveys used different approaches in determining the value of $v_{\rm micro}$. In APOGEE DR17 microturbulence is derived during the global fit of the spectrum \citep{abdurro_2021}. The microturbulence values published in GALAH DR3 are determined via the estimated empirical relations
\begin{equation*}
    v_{\rm micro} = 1.1 + 1.6\cdot10^{-4}\cdot(T_{\rm eff} - 5500~{\rm K})
\end{equation*}
if T$_{\rm eff}\leq$5500~K and $\log g \geq$4.2~dex or
\begin{equation*}
    v_{\rm micro} = 1.1 + 10^{-4}\cdot (T_{\rm eff} - 5500~{\rm K}) + 4\cdot10^{-7}\cdot (T_{\rm eff} - 5500~{\rm K})^2
\end{equation*}
otherwise \citep{buder_2021}.
According to \citet{buder_2021} the implementation of these empirical relations of $v_{\rm micro}$ provides more precise atomic abundance results. In addition,  several temperature-dependent trends (in the coolest and hottest regions) have been discovered by the GALAH team during the validation of element abundances. This issue can also be partially caused by over- or underestimated $v_{\rm micro}$ \citep{buder_2021}.

Furthermore, we examine the derived $v_{\rm micro}$ discrepancies between APOGEE  and GES. Since cuts beyond the global filtering are applied to opt out GES stars without any reported microturbulence data, we have 405 (mostly RGB) stars used in determining the mean difference of $-0.16$~km/s with a scatter of $0.38$~km/s. As in the case of APOGEE$-$GALAH, we recognize a growing linear trend in the $v_{\rm micro}-\Delta v_{\rm micro}$ plane (\autoref{matrix_param}). We note that the difference vanishes around 1.5~km/s and is systematically proportional to the $v_{\rm micro}$ data. 

As for the GALAH$-$GES catalog, we only take GES stars with available microturbulence data into account. Therefore, the statistics is made on 228 stars here. An overall mean offset of $-0.14$~km/s is observed with a scatter of $0.22$~km/s, which means a $-0.10\pm0.23$~km/s and a $-0.18\pm0.19$~km/s scatter deviation for dwarfs and giants, respectively (see \autoref{sum_apogalges2}). 
Due to the small number of GALAH$-$GES stars considered here, we find no relevant trends in $\Delta v_{\rm micro}$.

As for the analysis of microturbulent velocity, we find that APOGEE uses systematically lower $v_{\rm micro}$ values than GALAH and GES. We observe moderate discrepancy trends with respect to $T_{\rm eff}$, [M/H], and $\log g$, and a substantial correlation as a function of $v_{\rm micro}$.

\section{Discussion of differences: Chemical abundances} \label{disc_diff_abund}

In this section we describe the comparison of chemical abundance ratios, especially the combined $\alpha$ abundance and the individual species of elements that are present in the APOGEE$-$GALAH catalog (see the relevant intersections in \autoref{venn}). As discussed in Sect.~\ref{common_descr}, the  APOGEE$-$GALAH sample covers a wide range of effective temperature and surface gravity. Therefore, it is reasonable to provide a detailed further investigation on abundance discrepancy trends and correlations of the APOGEE$-$GALAH common data as it covers the largest parameter space and has the largest number of samples.

\begin{table}
\caption{Median absolute deviation of the reported abundances from the golden samples.}
\vspace{-.3cm}
\begin{tabular}{lp{0.15\textwidth}p{0.1\textwidth}p{0.1\textwidth}}
\hline \hline & 
\multicolumn{2}{c}{\textbf{APOGEE$-$GALAH}}
\\ \hline
 & \textbf{dwarfs} & \textbf{giants}  \\ \hline \hline 
$\Delta$[C/H] & 0.067 & ... \\
$\Delta$[O/H] & 0.119 & 0.162 \\
$\Delta$[Na/H] & 0.141 & 0.115 \\
$\Delta$[Mg/H] & 0.070 & 0.078 \\
$\Delta$[Al/H] & 0.109 & 0.074 \\
$\Delta$[Si/H] & 0.046 & 0.045 \\
$\Delta$[K/H] & 0.116 & 0.142 \\
$\Delta$[Ca/H] & 0.072 & 0.085 \\
$\Delta$[Ti/H] & 0.147 & 0.113 \\
$\Delta$[Ti~II/H] & 0.266 & 0.146 \\
$\Delta$[V/H] & 0.155 & 0.263 \\
$\Delta$[Cr/H] & 0.136 & 0.098 \\
$\Delta$[Mn/H] & 0.128 & 0.139 \\
$\Delta$[Co/H] & 0.240 & 0.102 \\
$\Delta$[Ni/H] & 0.082 & 0.081 \\
\hline \hline 
& \multicolumn{2}{c}{\textbf{APOGEE$-$GES}} \\
\hline 
& \textbf{dwarfs} & \textbf{giants} \\ 
\hline \hline  
$\Delta$[C/H] & {0.066} & {0.145} \\
$\Delta$[N/H] & {0.048} & {0.069} \\ 
$\Delta$[O/H] & {0.118} & {0.180} \\
$\Delta$[Na/H] & {0.177} & {0.142} \\
$\Delta$[Mg/H] & {0.057} & {0.065} \\
$\Delta$[Al/H] & {0.085} & {0.066} \\
$\Delta$[Si/H] & {0.072} & {0.049} \\
$\Delta$[S/H] & {0.115} & {0.333} \\
$\Delta$[Ca/H] & {0.080} & {0.101} \\
$\Delta$[Ti/H] & {0.143} & {0.112} \\
$\Delta$[Ti~II/H] & {0.315} & {0.104} \\
$\Delta$[V/H] & 0.106 & 0.181 \\
$\Delta$[Cr/H] & {0.126} & {0.100} \\
$\Delta$[Mn/H] & 0.131 & 0.124 \\
$\Delta$[Co/H] & 0.176 & 0.101 \\
$\Delta$[Ni/H] & {0.041} & {0.062} \\
\hline \hline 
& \multicolumn{2}{c}{\textbf{GALAH$-$GES}} \\
\hline 
& \textbf{dwarfs} & \textbf{giants} \\ 
\hline \hline 
$\Delta$[Li/H] & {0.167} & {0.128} \\ 
$\Delta$[C/H] & {0.079} & {0.087} \\
$\Delta$[O/H] & {0.114} & {0.134} \\
$\Delta$[Na/H] & {0.131} & {0.092} \\
$\Delta$[Mg/H] & {0.062} & {0.115} \\
$\Delta$[Al/H] & {0.072} & {0.068} \\
$\Delta$[Si/H] & {0.056} & {0.087} \\
$\Delta$[Ca/H] & {0.082} & {0.129} \\ 
$\Delta$[Sc/H] & 0.100 & 0.078 \\
$\Delta$[Ti/H] & {0.082} & {0.124} \\
$\Delta$[Ti~II/H] ~~~~~~~~~~~~& {0.096} & {0.166} \\
$\Delta$[V/H] & 0.120 & 0.178 \\
$\Delta$[Cr/H] & {0.097} & {0.090} \\
$\Delta$[Mn/H] & 0.067 & 0.159 \\
$\Delta$[Co/H] & 0.277 & 0.121 \\
$\Delta$[Ni/H] & {0.080} & {0.084} \\ 
$\Delta$[Cu/H] & 0.110 & 0.137 \\ 
$\Delta$[Zn/H] & 0.088 & 0.170 \\ 
$\Delta$[Zr/H] & {0.187} & {0.124} \\ 
$\Delta$[Mo/H] & ... & 0.111 \\ 
$\Delta$[Ba~II/H] & {0.133} & {0.274} \\ 
$\Delta$[La~II/H] & {0.252} & {0.186} \\ 
$\Delta$[Ce~II/H] & {0.115} & {0.162} \\
$\Delta$[Nd~II/H] & {0.264} & {0.126} \\ 
$\Delta$[Eu~II/H] & {0.019} & {0.156} \\ 
\hline
\label{mad_abund}
\end{tabular}
\end{table}

\begin{figure}
\centering
\includegraphics[width=.475\textwidth,angle=0]{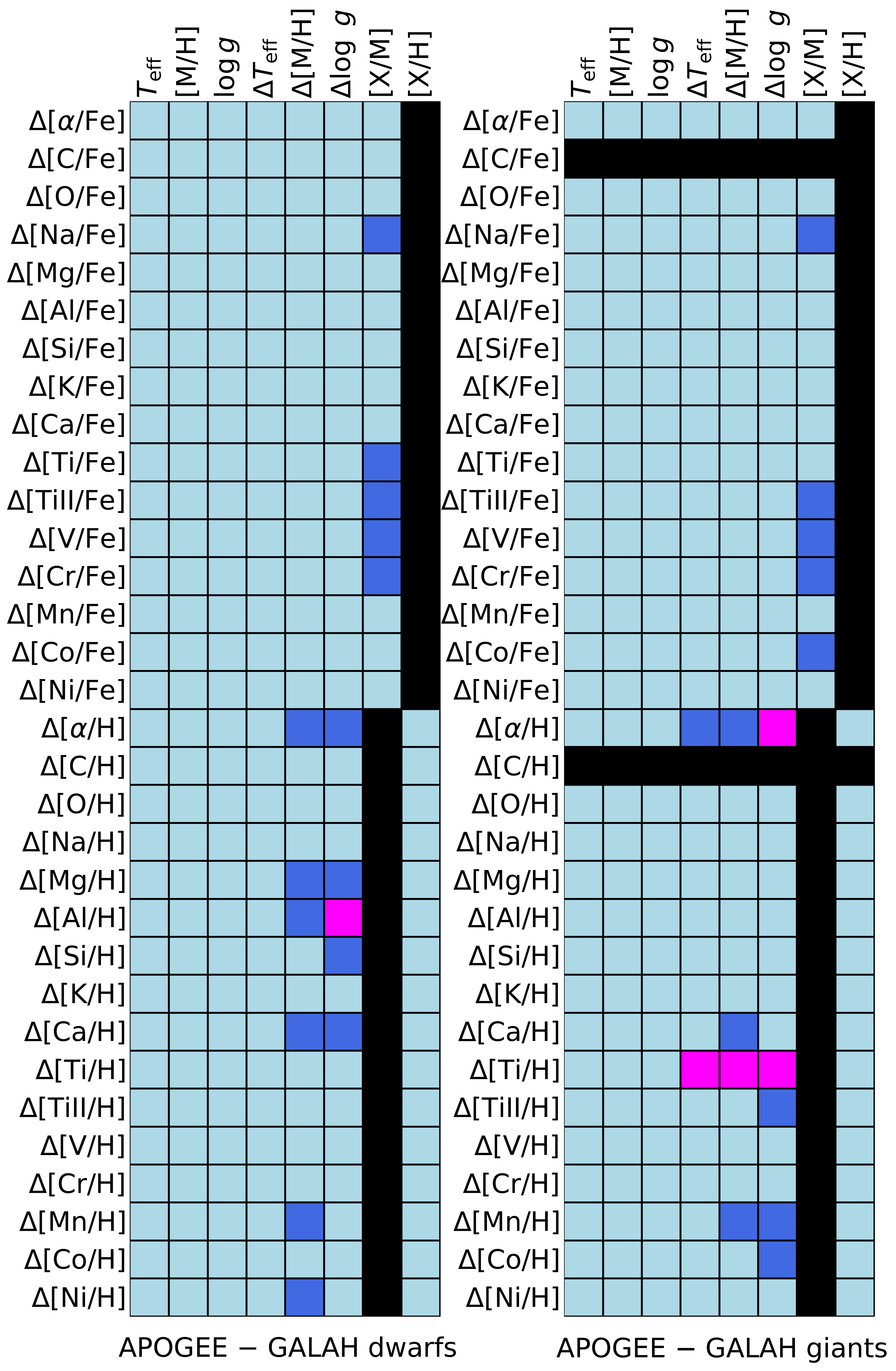}
\caption{Strength of correlations in the APOGEE$-$GALAH parameter and abundance pairs.  The color-coding is as follows:  Black cells are indeterminable, light blue cells represent no significant correlation ($<0.8\sigma$), dark blue cells indicate a weak correlation ($0.8\sigma-1.5\sigma$), and the magenta cells  mean strong correlation ($>1.5\sigma$). Our estimations   are based on the average correlation of reported individual uncertainties of stellar parameters and abundances. Dwarf (left panel) and giant stars (right panel) are separated.}
\label{matrix_abund} 
\end{figure}

\subsection{Abundances from APOGEE, GALAH, and GES}\label{6.1}

The APOGEE DR17 pipeline (ASPCAP, for details see Sect.~\ref{analysis} and \citealt{garcia_2016}, \citealt{jonsson_2020}) determines the 
chemical abundances based on   non-local thermodynamic equilibrium (NLTE) populations for Na, Mg, K, and Ca \citep{osorio_2020}, and LTE for the other elements. DR17 used the spectral synthesis code Synspec \citep{hubeny_2017,hubeny_2021}. Moreover, certain modifications to the linelists \citep{smith_2021} have been added \citep{abdurro_2021}.) 
We note that the estimated uncertainties of abundances are calculated by the APOGEE team using repeated target observations and result in a typical precision of $0.05$~dex \citep{abdurro_2021}.

As for GALAH, the abundance determination is also the second step of the parameter analysis of SME; the main parameters are kept fixed while iteratively fitting the individual abundances \citep{buder_2021}. We note that NLTE models are adopted for several elements common between APOGEE DR17 and GALAH DR3: C, O, Na, and Mg. In the final release, GALAH reports  relative abundances\footnote{The customary scale for logarithmic abundances is defined as $A(X)=\log(N_X/N_H) + A(H)$, where $A(H)\equiv12$ and $N_X$ is the number of atoms of element X per unit volume in the atmosphere.} [X/Fe] \citep{buder_2021}. Considering only the abundances from APOGEE$-$GALAH catalog, both line-by-line ($\alpha$, C, K, Ti, V, Co, Ni, Cu) and combined (O, Na, Mg, Al, Si, Ca, Cr, Mn) analyses were introduced in the GALAH pipeline. The combination of some elemental lines at different wavelengths is dependent on the detection of these lines and these estimates are a combination of 1 to 9 measurements. As described by \citet{buder_2021}, the Sun and solar twins were used in the zero-point validation of the abundance accuracy (further validations involved comparisons to Arcturus, GBS and cluster stars). Precision of individual abundances is assessed with a method based on the S/N scaled uncertainties of repeat observations and internal fitting covariance uncertainties from {SME} \citep{buder_2021}. For our calculations of absolute abundances, we make use of the $A(X)_{\odot}$ zero-points of Table A2 in \citet{buder_2021}.

\subsection{Comparison of surveys}\label{6.2}

We limit our discussion of comparisons with GES just to the absolute abundances due to the lack of enough common stars for a detailed analysis. Individual elemental abundance analysis uses local cuts (introduced in Sect.~\ref{data}) along with opting out stars with abundance values $\rm [X/H]<-1$ or $\rm [X/H]>1$. The mean and scatter of individual discrepancies for stars of the three common catalogs are shown in \autoref{sum_apogalges3}.

The number of species of elements reported by both APOGEE and GES is 17, for all of which the mean discrepancy ranges from
{$-0.257\pm0.319$~dex (in [Ti~II/H], for dwarfs) to $0.086\pm0.097$~dex (in [Ca/H], for dwarfs). }
The combined uncertainties exceed $\sigma_{\text{\tiny [X/H]}}\approx\sqrt{0.05^2+0.1^2}=0.11$~dex for the following elements: Na, S, Sc, Ti~II, V, Co (dwarfs); and  Si, S, Co (giants). This estimated precision is exceeded notably by the offsets from  {Ti~II ($\Delta \rm [Ti~II/H]=-0.257$) for the dwarfs and O ($\Delta \rm [O/H]=-0.202$) for the giants.} We note that the sample size is reduced by the local cuts to an extent that the final considered number of data points is usually too low for a more detailed analysis.

There are 29 elements measured in common between GALAH DR3 and GES DR5.  \autoref{sum_apogalges3} shows the average discrepancies with scatters. Ce~II (for giants) and {Co} (for dwarfs) have the most significant offsets among the listed elements:  {$-0.196\pm0.224$~dex and $0.384\pm0.467$~dex}. A precision of $0.1$~dex is estimated for both surveys for all elements, which results in a combined uncertainty of $\sigma_{\text{\tiny [X/H]}}\approx0.14$~dex, which do not cover the mean discrepancy range of the following species: Co, La~II, Nd~II, and Ce~II for dwarfs and giants (see \autoref{sum_apogalges3}). As  before, there are not enough stars in common to do a deeper analysis.

\begin{figure*}
\centering
\includegraphics[width=\textwidth,angle=0]{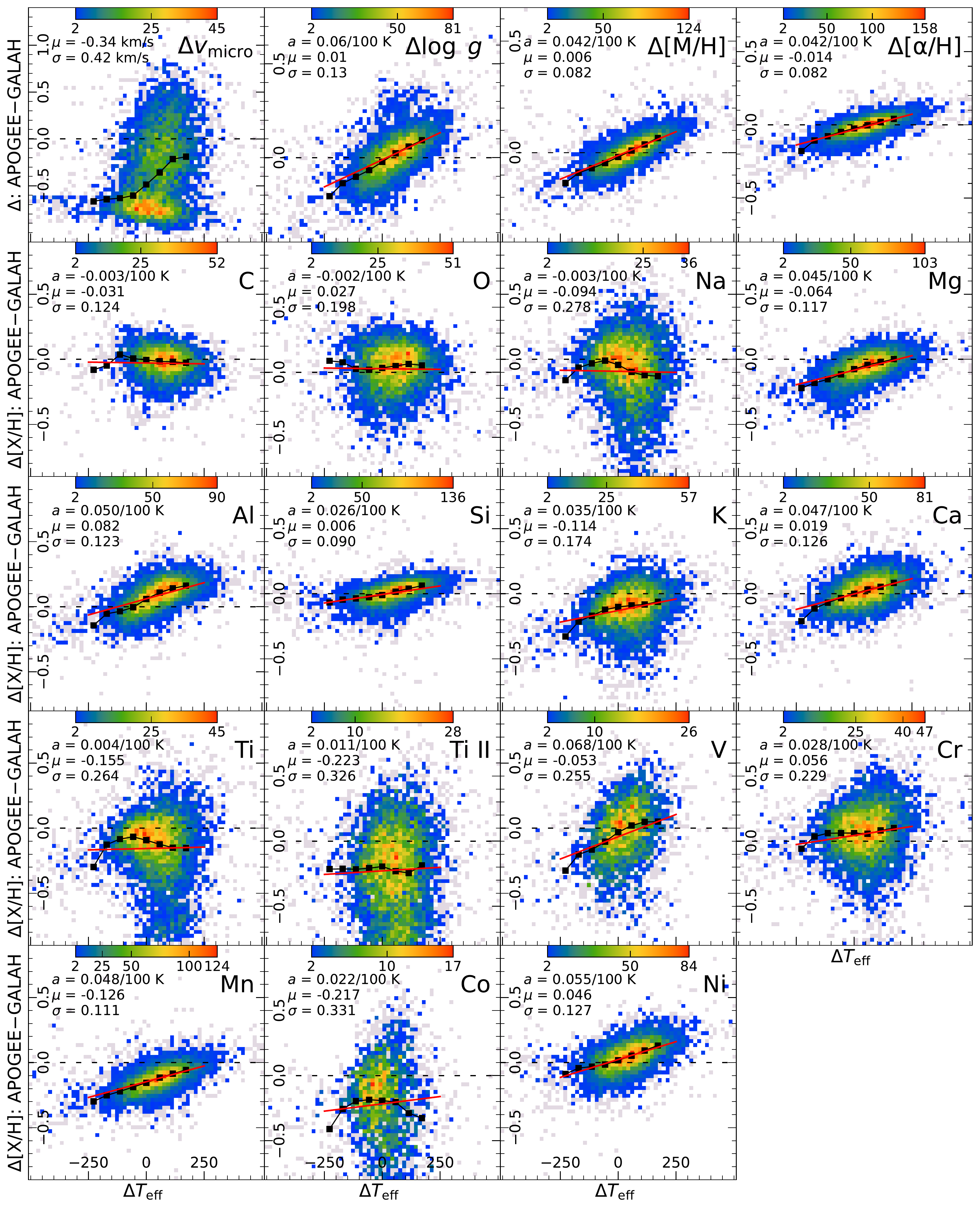}
\caption{Distributions of $\Delta v_{\rm micro}$ (km/s), $\Delta \log g$ (dex), $\Delta$[M/H], and $\Delta$[X/H] abundance discrepancies with respect to $\Delta  T_{\rm eff}$ for \textbf{MS stars} common between the APOGEE and GALAH quality data sets. The color-coding indicates the number of stars in the parameter planes, and bins representing one star are in gray. Black squares represent the median value within 50~K ranges. Red lines, with a slope of $a$, are fitted on the individual data points. The mean of the parameter differences ($\mu$) and the standard deviations ($\sigma$) are displayed in the top left corner of each panel.}
\label{dteff_dwarfs}
\vspace{1cm}
\end{figure*}

\begin{figure*}
\centering
\includegraphics[width=\textwidth,angle=0]{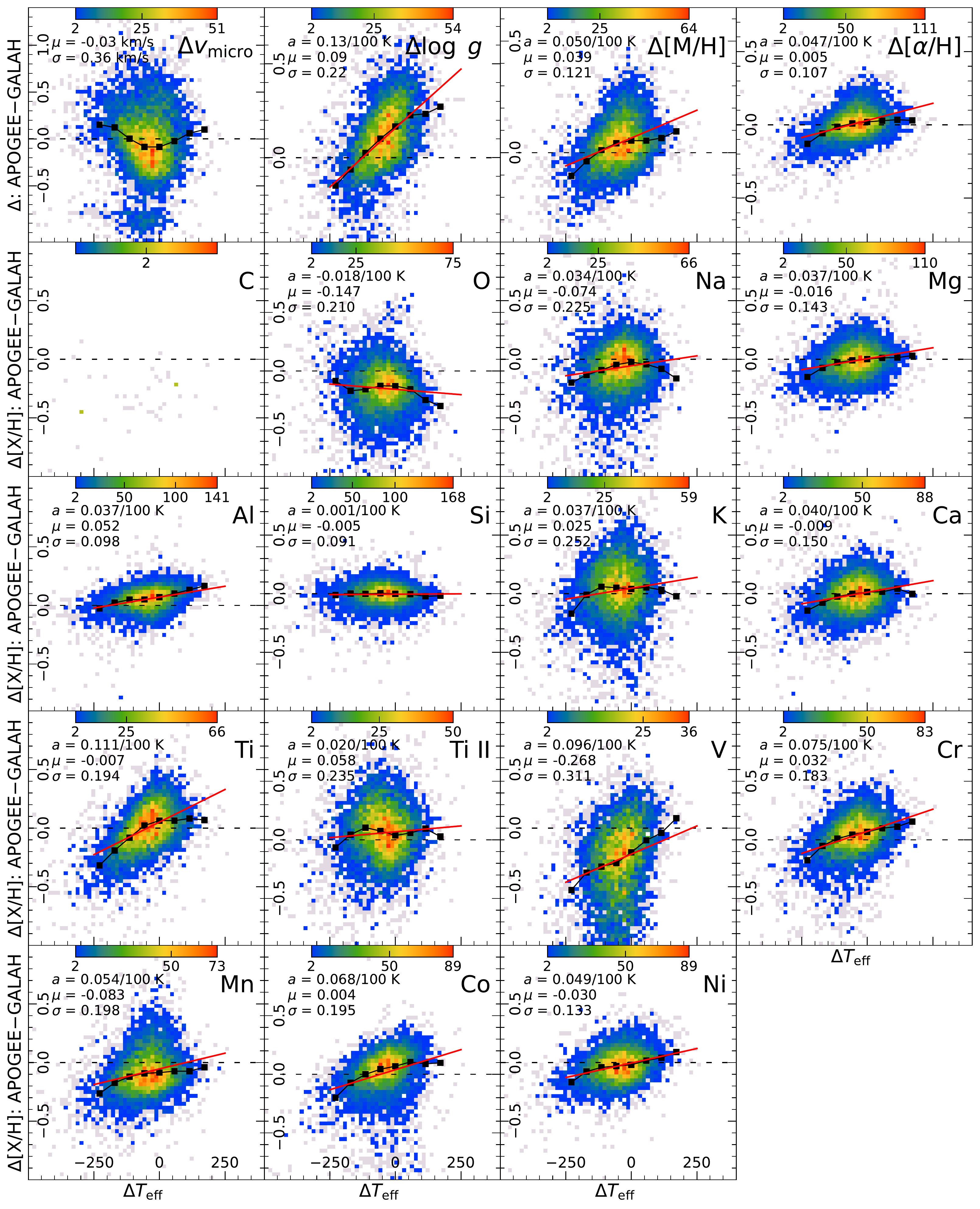}
\caption{Distributions of $\Delta v_{\rm micro}$ (km/s), $\Delta \log g$ (dex), $\Delta$[M/H], and $\Delta$[X/H] abundance discrepancies with respect to $\Delta T_{\rm eff}$ for \textbf{RGB stars} common between the APOGEE and GALAH quality data sets. The color-coding indicates the number of stars in the parameter planes, and bins representing one star are shown in gray. Black squares represent the median value within 50~K ranges. Red lines, with a slope of $a$, are fitted on the individual data points. The mean of the parameter differences ($\mu$) and the standard deviations ($\sigma$) are displayed in the top left corner of each panel.}
\label{dteff_giants}
\vspace{1cm}
\end{figure*}

\begin{figure*}
\centering
\includegraphics[width=\textwidth,angle=0]{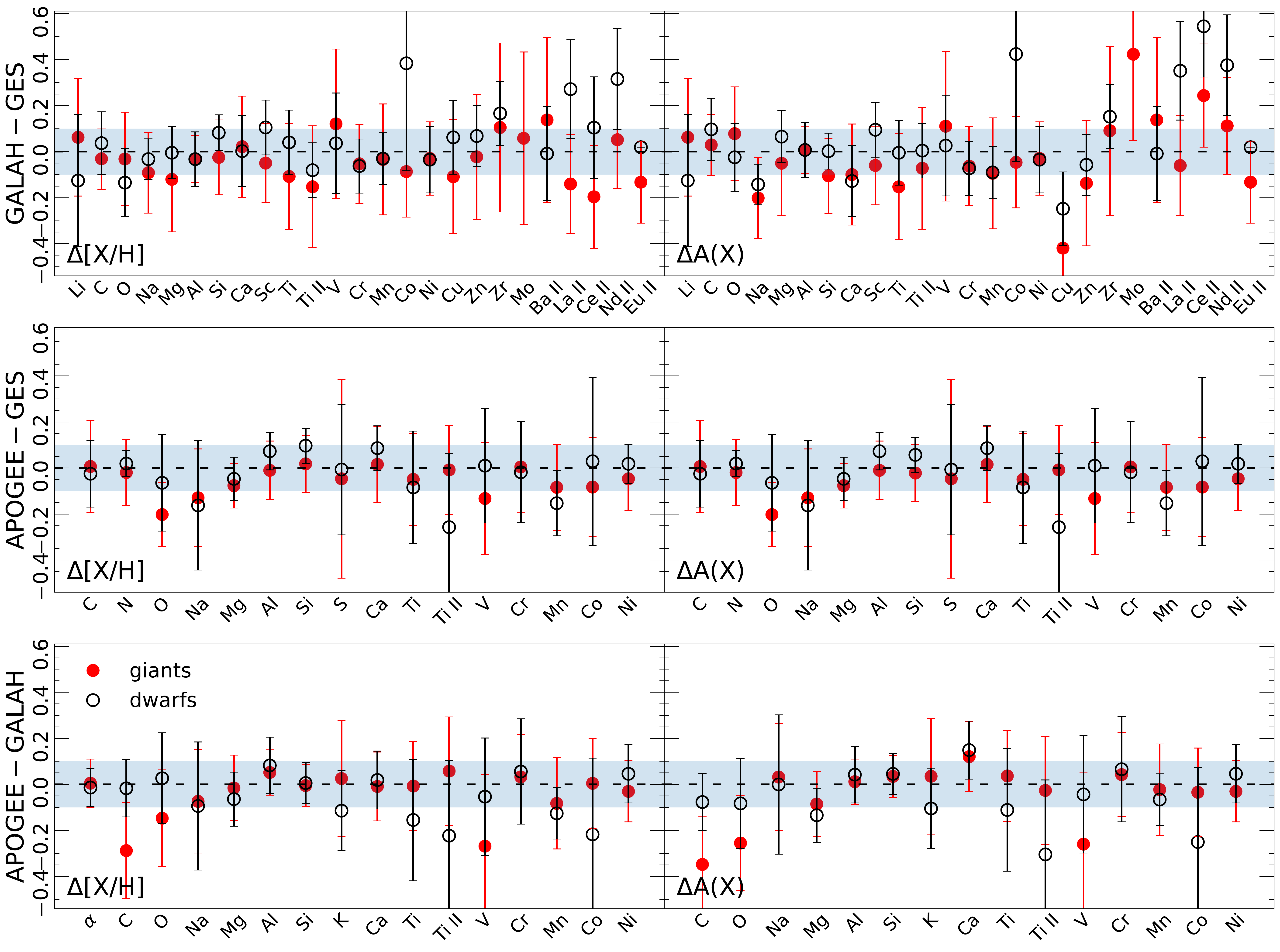}
\caption{Individual elemental abundance differences for APOGEE$-$GALAH (bottom), APOGEE$-$GES (middle), and GALAH$-$GES (top) common samples relative to hydrogen (left column) and on an absolute scale (right column). The shaded interval accounts for a $\pm 0.1$~dex range. The calculated results for GB and MS stars are shown as  filled red and empty black circles, respectively.}
\label{all_delta_x/h}
\end{figure*}

Below we give a detailed discussion about the differences and correlations for the APOGEE$-$GALAH common chemical abundances. \autoref{matrix_abund} shows the result of the analysis that is described here, and is analogous to \autoref{matrix_param}.
Each cell connects two quantities in a way that we fit a linear function on the abundance discrepancy, indicated by the row, with respect to the parameter indicated by the column. Considering the slope of the linear fit and knowing the expected precisions of the two relevant parameters and abundances, we deduce the significance of the correlations of uncertainties.
In \autoref{dteff_dwarfs}$-$\ref{dteff_giants}, we show the elemental distributions in the $\Delta{\rm (AP)}-\Delta$[X/H] planes, and possible interpretations are pointed out in this section. We note that samples represented in these plots have already met the previously described global and local filtering criteria. 
Moreover, \autoref{all_delta_x/h} shows the average and scatter of the elemental abundances contained in the three common catalogs relative to hydrogen (left column) and on an absolute scale where the zero-point offsets have been applied (right column).
We skip the analysis of [Cu/H] because \citet{abdurro_2021} consider the derivation of copper abundance in APOGEE DR17 to be unsuccessful, and this issue had also been recognized by us. We also leave out elements that have no available data after all the cuts.

\vspace{.4cm}
\noindent
\textbf{Alpha-process elements, \rm [$\alpha$/H]}.
The [$\alpha$/H] abundance is one of the independent parameters that ASPCAP uses in its simultaneous multidimensional fit for determining stellar parameters, and contains the following elements: O, Ne, Mg, Si, Ca, and Ti \citep{jonsson_2020}. 
For GALAH DR3, the reported alpha-enhancement [$\alpha$/Fe] is computed as the error-weighted combination of abundances of Mg (Mg5711), Si (combined), Ca (combined), and Ti (Ti4758, Ti4759, Ti4782, Ti4802, Ti4820, Ti5739) obtained by 1 to 9 measurements, depending on the detection of the lines; the relevant selected lines are written in brackets. As discussed in \citet{buder_2021}, the most dominant $\alpha$-element is Si, followed by Ti. 

The mean $\Delta\rm [\alpha/H]$ is $-0.005\pm0.096$~dex and there is good agreement between APOGEE DR17 and GALAH DR3. However, linear trends are present as functions of the parameter differences. According to \citet{jonsson_2020}, determinations by APOGEE in the near-infrared generally provide lower values of $\alpha$-abundance than measurements conducted in the optical ranges, which might be reflected in the mean overall discrepancy being below zero.

We find a mean difference of $-0.014\pm0.082$~dex in [$\alpha$/H] for the MS stars. We note that no systematic trend is found with respect to $\Delta v_{\rm micro}$ and the distribution is roughly symmetric with its low scatter. \autoref{matrix_abund} indicates a weak correlation with respect to $\Delta$[M/H] and $\Delta\log g$. In contrast, the offset is $0.005\pm0.107$~dex when considering RGB stars, with a decreasing trend in the $\Delta v_{\rm micro}-\Delta$[$\alpha$/H] distribution. Moreover, $\Delta T_{\rm eff}$ and $\Delta \rm [M/H]$ show weak correlations, while a strong correlation is found for $\Delta\log g$.

The main distinction for the $\alpha$-elements is that O is included in APOGEE,  while O is excluded in the GALAH $\alpha$-elements. However, since similar trends in the O comparison in \autoref{matrix_abund} are not seen, the exclusion of O within the GALAH procedure does not account for the full discrepancy seen (see the paragraph of O later).

\vspace{.4cm}
\noindent
\textbf{Carbon, \rm [C/H]}. 
Similarly to the case of $\alpha$-abundance, the [C/H] abundance determined by APOGEE is based on the CO molecular lines whose abundance is fitted simultaneously along with the stellar parameters and subsequently held fixed. We find good agreement between dwarfs in the APOGEE and GALAH quality samples, with a mean difference of $-0.031\pm0.124$~dex in the carbon abundance, and no correlations are found for $\Delta$[C/H] for the dwarf sample. Due to the lack of GALAH quality giants with responsible [C/H] data, results for giants are not found (see \autoref{sum_apogalges3}).
According to \autoref{all_delta_x/h}, the mean differences remain around the same value when considering absolute carbon abundances. 

\vspace{.4cm}
\noindent
\textbf{Oxygen, \rm [O/H]}. 
The oxygen abundance is determined with OH lines by the APOGEE pipeline. As  molecular lines become weak with increasing $T_{\rm eff}$, the O abundance tag is populated only for stars with $T_{\rm eff} < 5000~\rm K$. Oxygen shows a high precision and accuracy in the APOGEE data \citep{jonsson_2020}. No significant trend is seen with respect to the parameters (see Figures  \ref{matrix_abund}-\ref{dteff_giants}), but note that the average difference is $0.027$~dex with a scatter of $0.198$~dex and $-0.147$~dex with a scatter of $0.210$~dex for dwarfs and giants, respectively (see \autoref{sum_apogalges3}), meaning that GALAH systematically delivers higher [O/H] values for giants. 
The mean offsets become a little smaller by placing the O abundances on an absolute scale (see \autoref{all_delta_x/h}).  

\vspace{.4cm}
\noindent
\textbf{Sodium, \rm [Na/H]}. 
APOGEE spectra contain two weak lines of sodium, which may coincide with telluric lines as well; therefore, the reported precision is low \citep{jonsson_2020}. The final calibrated \texttt{NA\_FE} tag contains reliable data for giants, but deviant values for dwarfs (online documentation: abundances\footnote{\url{https://www.sdss.org/dr17/irspec/abundances/}}).
As for the dwarfs, we find an offset of $-0.094\pm0.278$~dex in $\Delta$[Na/H], with no recognizable trends. We note that numerous stars with $\Delta$[Na/H]~$<-1$ are removed by local cuts, and a significant scatter is still observed in \autoref{dteff_dwarfs}. For the giants there is a smaller offset and scatter: $-0.074$ and $0.225$~dex (\autoref{sum_apogalges3}). Giant stars display a growing trend in $\Delta$[Na/H] as a function of metallicity discrepancy, but a decreasing trend is observed with respect to $\Delta v_{\rm micro}$.
Moreover, we also find good agreement between APOGEE and GALAH on an absolute scale ($\Delta\rm A(Na)$).
Weak correlations are only found between $\Delta$[Na/Fe] and [Na/M] for both dwarfs and giants and no other sets of parameters (see \autoref{matrix_abund}).

\vspace{.4cm}
\noindent
\textbf{Magnesium, \rm [Mg/H]}. 
According to \citet{jonsson_2020}, Mg is one of the most precisely derived chemical elements in APOGEE, together with high accuracy, and is derived in DR17 assuming NLTE.
The analysis of magnesium abundance indicates a linear trend for dwarfs, similar to those shown for metallicity and $\alpha$-abundance in \autoref{dteff_giants} (as Mg contributes to both) with respect to $\Delta T_{\rm eff}$, $\Delta\log g$, and $\Delta$[M/H]. The mean difference is $-0.064\pm0.117$~dex. As \autoref{matrix_abund} indicates, there is a weak correlation with $\Delta\rm [M/H]$ and $\Delta\log g$. 
For the giants, no correlations are found (see \autoref{matrix_abund}), though a growing linear trend is present with respect to $\Delta$[M/H] (see \autoref{dteff_giants}), together with a mean difference of $-0.016\pm0.143$~dex (\autoref{sum_apogalges3}).
\autoref{all_delta_x/h} indicates good agreement both in [Mg/H] and A(Mg) between APOGEE and GALAH.

\vspace{.4cm}
\noindent
\textbf{Aluminum, \rm [Al/H]}. 
Three relatively strong Al lines are detected within the APOGEE spectral range, and  good precision is reported for the derived aluminum abundances both for giants and dwarfs. No correlations are found for $\Delta$[Al/Fe] (see \autoref{matrix_abund}). For $\Delta$[Al/H], a weak correlation is found between $\Delta$[M/H] and a strong correlation with $\Delta\log g$ for the dwarf sample, but none for the giants (as in \autoref{matrix_abund}). We find an offset of $0.082$~dex with a scatter of $0.123$~dex, and an increasing linear trend with respect to $\Delta T_{\rm eff}$ for dwarfs as seen in \autoref{dteff_dwarfs}. For giants there is slight trend with $\Delta T_{\rm eff}$ yielding a mean discrepancy of $0.052$~dex with a low scatter of $0.098$~dex (see \autoref{sum_apogalges3} and \autoref{dteff_dwarfs}, \ref{dteff_giants}), though the slope of the fitted trend remains lower than $0.8\sigma$, so there is no formal correlation.
We conclude that there is agreement of the aluminum abundances, both   relative to hydrogen and in an absolute way (see \autoref{all_delta_x/h}). Aluminum  was derived by assuming LTE by APOGEE and NLTE effects may introduce a  systematic offset as a function of $\Delta T_{\rm eff}$ and $\Delta \log g$ over certain parameter space.

\vspace{.4cm}
\noindent
\textbf{Silicon, \rm [Si/H]}. 
Silicon is reported to be one of the most precisely measured elements in APOGEE, and no significant trend was discovered in previous comparisons to optical measurements  \citep{jonsson_2020}.
Here we also find a negligible offset and standard deviation both for MS and RGB stars (\autoref{sum_apogalges3}, Figures \ref{dteff_dwarfs}--\ref{all_delta_x/h}). In addition to  the close overall agreement between APOGEE and GALAH, we note that a slight systematic difference is indicated in $\Delta\log g$ (also shown in \autoref{matrix_abund}) for dwarfs. 

\vspace{.4cm}
\noindent
\textbf{Potassium, \rm [K/H]}. \citet{jonsson_2020} reports a medium estimated precision of measured K abundance, which is based on two absorption lines. In DR17, NLTE calculations are introduced for K. 
In GALAH DR3, K is estimated from the $7699$~\AA~resonance line, which is also present in interstellar matter. Therefore, [K/H] should be used with caution in highly extinct regions \citep{buder_2021}.
As can be seen in \autoref{matrix_abund} potassium has no correlations for either dwarfs and giants, and [K/H] abundances suffer from a discrepancy of $-0.114\pm0.174$~dex  and $0.025\pm0.252$~dex between APOGEE and GALAH quality dwarf and giant stellar samples, respectively. With respect to metallicity, a growing systematic difference is displayed for both dwarfs and giants, suggesting that higher metallicity values convey higher measured abundances. We note that all these trends remain lower than $0.8\sigma$. Again NLTE effects may explain some of these discrepancies.

\vspace{.4cm}
\noindent
\textbf{Calcium, \rm [Ca/H]}. 
In APOGEE, the precision of Ca abundance is reported to be high, but a small offset may be required when compared to the optically determined  Ca abundance. Except for microturbulence, there appears to be a weak positive linear trend in the Ca abundance with increasing stellar parameters, but overall show fair agreement among the surveys. MS stars produce an average difference of $0.019$~dex with a $0.126$~dex scatter, while an offset of $-0.009$~dex and a scatter of $0.150$~dex is found in RGB stars. We note, however, that by placing the abundance on the absolute scale, the mean offsets become larger than $0.1$~dex. Some weak correlations in [Ca/H] are shown in \autoref{matrix_abund}, with $\Delta$[M/H] in both dwarfs and giants, and for $\Delta\log g$ in dwarfs.

\vspace{.4cm}
\noindent
\textbf{Titanium \rm I/II, \rm [Ti/H]/[TiII/H]}. 
As stated in \citet{jonsson_2020}, Ti~I in APOGEE shows low scatter if compared to optical data, but issues with some of the lines may occur, or the strong uncalibrated $T_{\rm eff}$ dependence of Ti~I lines propagate to individual abundance values. 
As there is only one Ti~II line in the APOGEE spectral window, its abundance is reported to have low precision and a significant scatter when compared to optical measurements \citep{jonsson_2020}. 
Since spectroscopic surface gravities are used in deriving the abundances, elements (such as Ti~II) that are sensitive to $\log g$ could have systematic offsets \citep{jonsson_2020}.

In the APOGEE$-$GALAH quality data set, Ti~I abundance values show a discrepancy of $-0.155\pm0.264$~dex and $-0.007\pm0.194$~dex for MS and RGB stars, respectively. We note that for giants the Ti~I abundance is derived more precisely than for dwarfs; however, the discovered trend is substantial especially in the $\Delta T_{\rm eff}-\Delta \rm [Ti/H]$, $\Delta\log g-\Delta \rm [Ti/H]$, and  $\Delta\rm [M/H]-\Delta \rm [Ti/H]$ planes that indicate strong correlations (see also \autoref{matrix_abund}). 
The case of ionized titanium is similar: there is an average difference of $-0.223\pm0.326$~dex and no clear trends among dwarfs, although the agreement is better ($0.058\pm0.235$~dex) for giants than for dwarfs, as also found by \citet{jonsson_2020}.
We also note   that the effective temperature dependence is weaker for $\Delta$[Ti~II/H] than for $\Delta$[Ti/H] (see \autoref{sum_apogalges3} and \autoref{dteff_dwarfs}, \ref{dteff_giants}).

\vspace{.4cm}
\noindent
\textbf{Vanadium, \rm [V/H]}. 
For APOGEE, the derived vanadium abundance is reported to have rather low precision, accuracy, and reliability by the online documentation (abundances), and it is advised to use the  V abundance with caution. Similarly, \citet{buder_2021} also gives a warning about the use of the  V abundance, as blended lines result in possible incorrect systematics. A weak  correlation is found between [V/M] and $\Delta$[V/Fe] in our comparison for both dwarfs and giants. For the dwarfs, there is a mean [V/H] difference of $-0.053$~dex and a $0.255$~dex scatter. However, the giants show a $-0.268\pm0.311$~dex discrepancy of [V/H] on average. For the absolute abundance A(V), we observe almost the same offsets (see \autoref{all_delta_x/h}).

\vspace{.4cm}
\noindent
\textbf{Chromium, \rm [Cr/H]}. 
Comparison with optical abundances showed that in APOGEE, chromium is measured with moderate precision and accuracy (online documentation: abundances). For the MS stars, a discrepancy of $0.056\pm0.229$~dex is present in $\Delta$[Cr/H], together with no weak systematic correlations with the parameters. The giants show an offset of $0.032$~dex and a scatter of $0.183$~dex in $\Delta$[Cr/H] and A(Cr) as well, and we observe various slight systematic trends with respect to all atmospheric parameters, which all have slopes lower than $0.8\sigma$.

\vspace{.4cm}
\noindent
\textbf{Manganese, \rm [Mn/H]}. 
Zero-point Mn abundance shifts are included in the final APOGEE release, although reporting a high precision and reliability. 
We establish weak correlations with $\Delta$[M/H] for both Ms and RGB stars, and with $\Delta \log g$ for the giants only a seen in \autoref{matrix_abund}. We find a linear $\Delta T_{\rm eff}$ trend with an offset of $-0.126$~dex and a low scatter of $0.111$~dex for the dwarfs (\autoref{dteff_dwarfs}), and an offset of $-0.083$~dex with a scatter $0.198$~dex for the giants (\autoref{dteff_giants}). In addition to the offset, the main parameter correlations mostly originate from the metallicity trends (see also \autoref{matrix_abund}). We note that by calculating the offsets on an absolute scale, the mean differences decrease to $<0.1$~dex.

\vspace{.4cm}
\noindent
\textbf{Cobalt, \rm [Co/H]}. 
APOGEE spectral windows involve one single cobalt line from which to derive its abundance, and thus a significant scatter is expected, especially among dwarfs. 
For the dwarfs we   find a substantial offset and scatter of $-0.217\pm0.331$~dex in Co, with no identified systematic pattern. However, $\Delta$[Co/H] shows a slight correlation, but lower than $0.8\sigma$,  similar to $\Delta$[M/H] with respect to the effective temperature differences for giants (see \autoref{dteff_giants}), but the mean discrepancy is lower: $0.004\pm0.195$~dex. As indicated in \autoref{matrix_abund}, there is a weak correlation in $\Delta$[Co/H] with $\Delta \log g$ for giants.

\vspace{.4cm}
\noindent
\textbf{Nickel, \rm [Ni/H]}. 
Nickel is derived with a high precision in APOGEE (online documentation: abundances), although some systematic deviations compared to optical studies have been revealed. Nickel is claimed to have the most accurate and precise elemental abundance among the iron-peak elements in APOGEE \citep{jonsson_2020}.
In $\Delta$[Ni/H], a clear linear trend is observed with respect to effective temperature, and the offset is $0.046\pm0.127$~dex for MS stars (\autoref{dteff_dwarfs}), while the giants have an offset of $-0.030\pm0.133$~dex (\autoref{dteff_giants}). For the dwarfs a systematic trend in $\Delta$[M/H] is seen in \autoref{matrix_abund}.
When considering absolute abundances the mean offsets lie within $0.1$~dex, indicating a general agreement.

\section{Effect on the Milky Way chemical maps}\label{mw_maps}

Comparison analyses like those presented in this work not only provide a reference basis for future stellar spectroscopy, but also reveal the effect of discrepancies on existing Milky Way chemical maps. Using data from APOGEE DR14 and GALAH DR2, \citet{griffith_2019} discovered that the GALAH and APOGEE median trends for Si and Ca are similar, but O has a much stronger metallicity dependence in GALAH, while the APOGEE trends are nearly flat.

\begin{figure}
\centering
\includegraphics[width=.45\textwidth,angle=0]{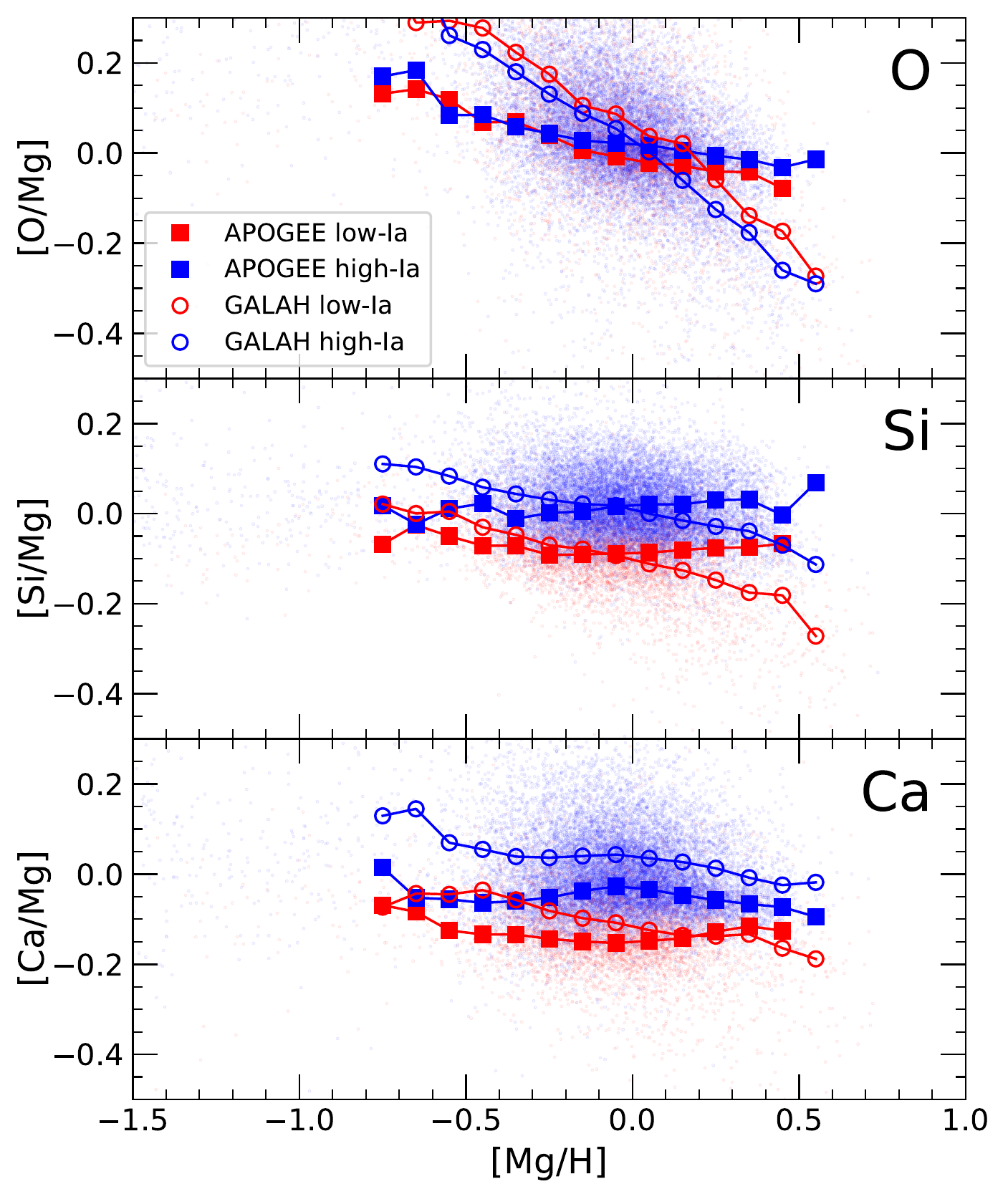}
\caption{Reproduction of Figure~3 from \citet{griffith_2019} based on the most current versions of the databases. GALAH DR3 $\alpha$-element median abundances of the high-Ia supernova (empty blue circles) and low-Ia supernova (empty red circles) populations. Data are binned by $0.1$~dex in [Mg/H]. APOGEE DR17 median abundances are also included (squares).}
\label{mw_map}
\end{figure}

\begin{figure*}
\centering
\includegraphics[width=\textwidth,angle=0]{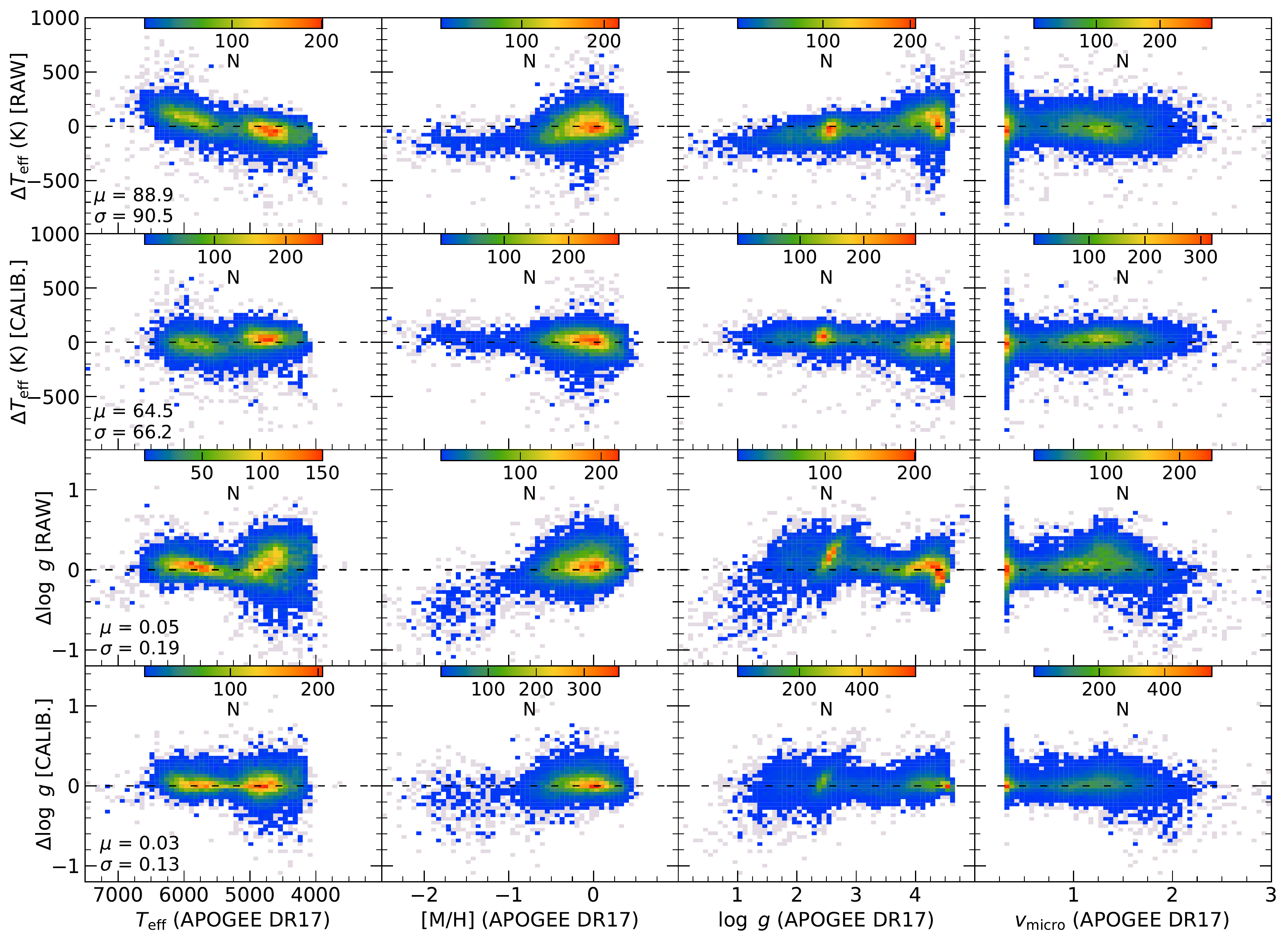}
\caption{Main parameters vs. $T_{\rm eff}$ and $\log g$ discrepancies based on raw (first and third rows) and calibrated (second and fourth rows) values from the APOGEE$-$GALAH common data set. Horizontal axes indicate APOGEE DR17 data in all panels. Every data point is marked in the parameter planes, while the color-coding of stellar density starts from number two. The mean and standard deviation of the differences are also shown in each case.}
\label{cal}
\end{figure*}

\autoref{mw_map} represents the [O/Mg], [Si/Mg], and [Ca/Mg] median trends versus [Mg/H] for the low-Ia (high-$\alpha$) and high-Ia (low-$\alpha$) Type Ia supernovae stellar populations of the APOGEE$-$GALAH filtered data set (additional criteria applied: $4500~{\rm K} < T_{\rm eff} < 6200~{\rm K}$ and the absence of relevant elemental flags), using the same definitions as  \citet{weinberg_2019} for the ``two-process model.'' In the updated version of Figure~3 in \citet{griffith_2019} and in the current paper we reveal that similar negatively sloped median trends are still present, as in APOGEE DR14, and also a stronger correlation in GALAH DR2.
The top panel of \autoref{mw_map} indicates close agreement of the low-Ia and high-Ia sequences. As  is theoretically known,  O and Mg yields are both dominated by core collapse supernovae. 
The disagreements in the slopes of the [M/H] or [Mg/H] dependences could arise from the difficulty of deriving oxygen abundances from both optical (GALAH, atomic lines) and near-IR (APOGEE, molecular lines) spectra \citep{weinberg_2019}.
The flat trend drawn by APOGEE data agrees better with theoretical expectations, as the O and Mg enrichment show a  slight dependence on metallicity.  Following \citet{weinberg_2019} and \citet{griffith_2019}, \citet{weinberg_2021} used Mg as a reference and  also fit the median trends of low-Ia and high-Ia populations and approximate stellar abundance patterns as the sum of a CCSN contribution and an SNIa contribution. 
Sharper separations between the low and high sequences are observed in the median trends of Si and Ca in the middle and bottom panels, respectively, suggesting that both elements have a non-negligible contribution from SNIa. 

In general, it would be crucial to understand and correct for the trends arising in O, as investigating the chemical evolution of the Galaxy requires reference elements produced and scattered dominantly by CCSN explosions (e.g., Mg and O).

\section{Effect of APOGEE DR17 calibration}\label{calib}

From APOGEE DR17 only raw parameter and chemical abundance values are used throughout our analysis, and this section includes the investigation of the effect of the APOGEE DR17 surface gravity and effective temperature calibrations on the previously seen discrepancy distributions. For the APOGEE$-$GALAH quality stars, \autoref{cal} shows the discrepancies in $T_{\rm eff}$ and $\log g$ before and after the APOGEE DR17 calibrations. Calibrated values come from the \texttt{PARAM} array, and the horizontal axes represent the raw main parameter values derived by APOGEE.

Calibrated $T_{\rm eff}$ is set by a comparison to photometric effective temperatures \citep{jonsson_2020}.
Unlike in DR16 \citep{jonsson_2020}, the $\log g$ calibration in DR17 was applied using a neural network,  and this new approach removes small discontinuities that were apparent (online documentation: parameters\footnote{\url{https://www.sdss.org/dr17/irspec/parameters/}}). 

By comparing the raw and corrected data in \autoref{cal}, the conclusion is clear. The implementation of the calibration processes results in a smaller offset with a scatter, and each substantial trend is eliminated with respect to the main parameters; therefore, the stellar density distributions show nearly symmetrical patterns. However, it is arguable whether the solar sample of calibrators used in the photometric effective temperature calibration represents the truth, and thus justifies those calibrations.

\section{Conclusion}\label{conclusion}

Comparing abundances between high-resolution surveys will become more and more important as new surveys such as MWM, 4MOST, and WEAVE will observe millions of stars and will be able to map the chemical makeup of our Galaxy in more detail than ever before. Revealing variants in the results between the surveys yields insight into the underlying physics and methodologies employed and informs future surveys. The work presented here contributes to the preparation for the abundances of the next generation of surveys, particularly the MWM. Similarly to APOGEE, MWM also uses SDSS spectrographs, and therefore  adds another reason why APOGEE is the main reference throughout the paper in addition to   the more extensive detailed coverage of the HRD.

APOGEE, GALAH, and Gaia-ESO are the first three major high-resolution spectroscopic sky survey programs operating over the last decade and providing publicly available data. In this work we created the APOGEE$-$GALAH, APOGEE$-$GES, and GALAH$-$GES stellar catalogs containing golden samples using their latest data releases.

Quality stars are collected according to a consistent quality criteria: sufficient S/N, low $v\sin i$, low scatter, and error of RV, and for technical issues the parameters and chemical abundances also utilize the flagging systems of the three surveys. The filtered common data set covers $4000{\rm~K}<T_{\rm eff}<7000{\rm~K}$ and $0{\rm~dex}<\log~g<4.75{\rm~dex}$ ranges in the HRD, and the metallicity spans $-1.5$~dex to $0.5$~dex.

We investigated the discrepancies and potential correlations of the RV, primary stellar parameters and common species of elemental abundances (both on relative and absolute scales) between the surveys. The analysis is based on comparing average differences and the slope of correlations with the reported individual uncertainties from APOGEE, GALAH, and GES. In most cases the median absolute deviation (MAD) of the differences is lower than the standard deviation of the differences both in the parameters and in the abundances, which indicates that in addition to  the offset and scatter there are no unexpected dispersion issues in the parameter planes.

Radial velocity analysis reveals that the mean of the differences is below the combined uncertainty intervals for APOGEE$-$GALAH, APOGEE$-$GES, and GALAH$-$GES quality stars. By separating the MS and RGB we find similarly good agreements between the surveys; however, significant scatters are still observed in RVs. The origin of this variation in $\Delta v_{\rm rad}$ that range from $1.26$~km/s (APOGEE$-$GES dwarfs) to $4.37$~km/s (GALAH$-$GES dwarfs) is still unclear, even after eliminating variable stars through cross-matching with the GCVS, with the APOGEE DR17 VAC of spectroscopic binaries and with the Gaia Variable Summary table. The presence of as-yet-unknown measurement discrepancies or undiscovered variables might explain this issue.

Discrepancies and possible correlations of effective temperature, surface gravity, metallicity, and microturbulence are discussed. According to the mean differences, all the main stellar parameters are generally measured consistently by the surveys, although we find possible weak correlations in the  parameter-planes $\Delta T_{\rm eff}$ versus $\Delta\rm[M/H]$ (for the APOGEE$-$GALAH dwarfs), $\Delta T_{\rm eff}$ versus $\Delta \log~g$ (for the APOGEE$-$GALAH dwarfs and APOGEE$-$GES stars), and in the $\Delta$[M/H] versus $\Delta \log~g$ (for APOGEE$-$GALAH stars).
Moreover, there is a clear growing linear trend in $\Delta v_{\rm micro}$ as a function of $v_{\rm micro}$, probably caused by the fact that APOGEE determines it as part of the global spectral fit, while GALAH calculates it from an empirical relation of $T_{\rm eff}$.

We discuss the mean and standard deviation of the common individual abundances for all three overlapping quality stellar catalogs; in addition, the strength of the correlations between species from APOGEE$-$GALAH and the main parameters is investigated. 
Generally, we draw the  conclusion that {the vast majority} of the common chemical abundance ratios are consistent in the relative scale. The strength of correlations are represented in \autoref{matrix_abund}, and, for example,  [Al/H] differences do have a strong linear dependence on $\Delta \log g$. The comparison of the relative and absolute abundance values (see \autoref{all_delta_x/h}) suggests that implementing the solar reference values slightly improves the overall agreement between the elements, but there are still large variations within the reference values used by the surveys. Although most of the differences are within the uncertainties, evaluating and understanding the sources of the error remains crucial to put Milky Way stellar data on the same scale as the new data and methods increasingly improve the precision.

\begin{acknowledgements}

VH and SzM acknowledge the support of Hungarian Academy of Sciences through MTA-ELTE Lend{\"u}let ``Momentum'' Milky Way Research Group.

Funding for the Sloan Digital Sky 
Survey IV has been provided by the 
Alfred P. Sloan Foundation, the U.S. 
Department of Energy Office of 
Science, and the Participating 
Institutions. 

SDSS-IV acknowledges support and 
resources from the Center for High 
Performance Computing  at the 
University of Utah. The SDSS 
website is www.sdss.org.

SDSS-IV is managed by the 
Astrophysical Research Consortium 
for the Participating Institutions 
of the SDSS Collaboration including 
the Brazilian Participation Group, 
the Carnegie Institution for Science, 
Carnegie Mellon University, Center for 
Astrophysics | Harvard \& 
Smithsonian, the Chilean Participation 
Group, the French Participation Group, 
Instituto de Astrof\'isica de 
Canarias, The Johns Hopkins 
University, Kavli Institute for the 
Physics and Mathematics of the 
Universe (IPMU) / University of 
Tokyo, the Korean Participation Group, 
Lawrence Berkeley National Laboratory, 
Leibniz Institut f\"ur Astrophysik 
Potsdam (AIP),  Max-Planck-Institut 
f\"ur Astronomie (MPIA Heidelberg), 
Max-Planck-Institut f\"ur 
Astrophysik (MPA Garching), 
Max-Planck-Institut f\"ur 
Extraterrestrische Physik (MPE), 
National Astronomical Observatories of 
China, New Mexico State University, 
New York University, University of 
Notre Dame, Observat\'ario 
Nacional / MCTI, The Ohio State 
University, Pennsylvania State 
University, Shanghai 
Astronomical Observatory, United 
Kingdom Participation Group, 
Universidad Nacional Aut\'onoma 
de M\'exico, University of Arizona, 
University of Colorado Boulder, 
University of Oxford, University of 
Portsmouth, University of Utah, 
University of Virginia, University 
of Washington, University of 
Wisconsin, Vanderbilt University, 
and Yale University.

{PJ acknowledges financial support of FONDECYT Regular Grant Number 1200703 and Millenium Nucleus  ERIS Number NCN2021\_017.}
\end{acknowledgements}

\bibliographystyle{aa}
\bibliography{references} 
\end{document}